\documentclass[10pt]{article}
\usepackage{graphicx}
\usepackage{float}
\usepackage[table]{xcolor}
\usepackage{longtable} 
\usepackage{color}
\usepackage{etoolbox}     
\usepackage{orcidlink}
\usepackage{hyperref}
\hypersetup{colorlinks=true, linkcolor=blue, citecolor=blue, urlcolor=blue}
\usepackage{subcaption}
\usepackage{amsmath}
\providecommand{\keywords}[1]
{
  \small	
  \textbf{\textit{Keywords---}} #1
}
\RequirePackage[numbers,sort&compress]{natbib}
\begin{document}
\baselineskip=16pt

\begin{center}
\large{\bf \color{black}{Geodesics, Scalar Fields, and GUP-Corrected Thermodynamics of Charged BTZ-like Black Holes in Bopp-Podolsky Electrodynamics}}    
\end{center}

\vspace{0.3cm}

\begin{center}

{\bf Faizuddin Ahmed}\orcidlink{0000-0003-2196-9622}\\
Department of Physics, Royal Global University, Guwahati, 781035, Assam, India\\
\textbf{E-mail}: faizuddinahmed15@gmail.com\\

\vspace{0.15cm}
{\bf Ahmad Al-Badawi}\orcidlink{0000-0002-3127-3453}\\
Department of Physics, Al-Hussein Bin Talal University, 71111,
Ma'an, Jordan \\
\textbf{E-mail}: ahmadbadawi@ahu.edu.jo\\
\vspace{0.15cm}
{\bf{Abdelmalek Bouzenada\orcidlink{0000-0002-3363-980X}}}\\ \textbf{E-mail}: abdelmalekbouzenada@gmail.com   \\
	Laboratory of Theoretical and Applied Physics, Echahid Cheikh Larbi Tebessi University 12001, Algeria.

    {\bf Erdem Sucu}\orcidlink{0009-0000-3619-1492} and {\bf \.{I}zzet Sakall{\i}}\orcidlink{0000-0001-7827-9476}\\
Physics Department, Eastern Mediterranean University, Famagusta 99628, North Cyprus via Mersin 10, Turkey\\
\textbf{E-mail}: erdemsc07@gmail.com;\quad\quad \textbf{E-mail}: izzet.sakalli@emu.edu.tr\\


\vspace{0.15cm}

\end{center}

\vspace{0.15cm}

\begin{abstract}
{\color{black}In Ref. \cite{EPJC}, the spacetime geometry generated by compact objects in (2+1)-dimensional Bopp-Podolsky electrodynamics is derived. Using a perturbative approach, the authors derived a charged BTZ-like black hole solution and computed corrections up to second order in a perturbative expansion valid far from the horizon. In this work, we investigate the same circularly symmetric three-dimensional charged BH solution pierced by disclinations. We begin by analyzing the motion of photons within this spacetime, focusing on how the geometric parameters influence the effective potential governing null geodesic motion. The corresponding equations of motion are derived, and the resulting orbital dynamics are explored through graphical methods to show the influence of key parameters including the BH mass $M$, electric charge $Q$, cosmological constant $\Lambda$, and BP coupling parameter $b^2$. Extending our analysis to wave dynamics, we examine the propagation of massless scalar fields in the BH solution by solving the Klein-Gordon equation. Through suitable coordinate transformations, we derive a Schrödinger-like equation with an effective potential that encodes the influence of the topological defect and electromagnetic corrections. Additionally, we investigate quantum gravitational effects by applying the Generalized Uncertainty Principle (GUP) to derive a GUP-corrected Hawking temperature, revealing systematic suppression of thermal radiation that could lead to stable BH remnants. Finally, we compute Keplerian frequencies for circular orbits, demonstrating how the interplay between charge, nonlinear electrodynamics, and disclination parameters creates distinctive observational signatures that could potentially test modified gravity theories in strong-field regimes.}
\end{abstract}


{\color{black}
\section{\mdseries{Introduction}}\label{sec1}
General Relativity (GR) not only theoretically predicts the existence of black hole (BH) models but also shows and illustrates to astronomers concrete methods to detect and study them through their gravitational influence on nearby electromagnetic radiation, another case of gravitational lensing where light rays are deflected from their original paths when passing through strong gravitational fields. In this case, the recent advent of horizon-scale observations of supermassive BH models by advanced instruments like the Event Horizon Telescope (EHT), tested in many studies \cite{Doeleman2008,Akiyama2019a,Akiyama2019b,Akiyama2019c,isz01,isz02}, has ushered in a new era of astrophysical exploration, enabling scientists to probe the intricate details of photon trajectories in the extreme spacetime curvature near BH event horizons \cite{Carter1968,Gralla2020}. The study of these geodesics provides critical information for understanding modern astronomy. While the initial theoretical models for gravitational lensing were developed under the weak-field approximation, assuming small deflection angles and thin lenses \cite{Liebes1964,Refsdal1964,Bourassa1975,Schneider1992}, the extreme conditions near BHs, where photons may orbit multiple times before escaping or falling into the event horizon, necessitate a complete strong-field treatment of the lensing phenomenon. Also, to illustrate the strong gravitational lensing, Darwin \cite{Darwin1959}, in his pioneering work, followed by significant contributions from Virbhadra and Ellis \cite{Virbhadra2000}, suggests and shows numerical computations of light deflection angles for Schwarzschild BHs in asymptotically flat spacetimes, providing crucial quantitative insights into how light behaves in such extreme environments. Also, theoretical models occurred when Frittelli, Kling, and Newman \cite{Frittelli2000} developed an exact lens equation through rigorous mathematical formulation that moved beyond previous approximations, while the most comprehensive advancement came from Bozza et al. \cite{Bozza2001}, who systematically investigated light bending in the strong-gravity regime near compact objects, deriving analytical expressions that could predict the formation, positions, and magnifications of multiple images when distant light sources are gravitationally lensed by black holes based on precise source-lens-observer geometries. These theoretical developments have proven indispensable for interpreting cutting-edge astronomical observations, testing the distinctive photon rings and shadow features revealed by recent horizon-scale BH imaging, while also testing GR predictions in extreme gravity regimes and potentially showing deviations that might indicate new physics, with future progress expected from synergies between gravitational wave astronomy, next-generation telescopes, and refined theoretical models that will continue to deepen our understanding of spacetime's behavior under the most intense gravitational conditions described by Einstein's theory \cite{Doeleman2008,Akiyama2019a,Akiyama2019b,Akiyama2019c,Carter1968,Gralla2020,Liebes1964,Refsdal1964,Bourassa1975,Schneider1992,Darwin1959,Virbhadra2000,Frittelli2000,Bozza2001}.

One of the fundamental predictions of Einstein's GR is the formation of black holes (BHs), representing regions of extreme curvature and causal disconnection from the rest of spacetime. Also, the earliest analytical model for such an object was obtained by the Schwarzschild model in 1916, but it was Finkelstein who later explained and suggested a physical interpretation of the event horizon as a unidirectional membrane in spacetime \cite{Finkelstein1958}. These results stimulated profound interest in the classification and investigation of exact BH model solutions. A milestone was reached in 1992, when Bañados, Teitelboim, and Zanelli (BTZ) formulated a three-dimensional ($1+2$) solution with a negative cosmological constant, now known as the BTZ BH \cite{Banados1992}. This solution became a fertile ground for studying fundamental questions related to quantum gravity, AdS/CFT duality, and topological aspects of spacetime \cite{Banados1993, Nojiri1998, Emparan2000, Hemming2002, Sahoo2006, Cadoni2008, Park2008, Parsons2009, Akbar2011, Myung2011, Hodgkinson2012, Moon2012, Frodden2013, Eune2013, Lemos2014, BravoGaete2014, ESUCU2025, Mansoori2016, Wu2016, Witten2007, Carlip2005,isz03,isz04}. BTZ BH solutions have been illustrated and generalized in the context of various gauge field theories, including linear and nonlinear electrodynamics such as Maxwell, power-Maxwell, and Born–Infeld theories \cite{Gurtug2012, Hendi2014, Chougule:2018cny, Cataldo1999, Myung2008, Mazharimousavi2015, Hendi2015, Tang2017, Yamazaki2001, Hendi2012}. Also, these geometries have been embedded into frameworks such as massive gravity, gravity's rainbow, and dilaton gravity \cite{Hendi2016a, Hendi2016b, Hendi2017b, Chan1994, Dehghani:2023yuq, Hendi2011, Hendi2012b}. In this context, BTZ BHs have been tested to include mass terms in modified gravity (MG), leading to charged ($Q$) and asymptotically AdS BH model solutions in massive gravity \cite{Hendi2016c}. These studies have illustrated examinations of their thermodynamic properties, including the use of geometric thermodynamics via Weinhold, Ruppeiner, and Quevedo formalisms. Also, the Hawking radiation (HR) of such BHs has been tested through the tunneling perspective, emphasizing its quantum mechanical nature \cite{Hawking1975, Parikh2000, Kim2011, Yale2011, nozari2008,isz05,isz06}. In the paradigm of BH chemistry, where the cosmological constant is reinterpreted as a pressure term, BTZ geometries offer a platform to investigate thermodynamic cycles, including BH heat engines \cite{Frassino2015, Kubiznak2017, Hennigar2017, Johnson2014}. These insights establish lower-dimensional ($\mathcal{D}$) BHs as essential laboratories for showing and illustrating the interface of gravitation, thermodynamics, and quantum field theory.

Recent investigations have demonstrated that BHs can exhibit event horizons with varied topologies, allowing their formation through gravitational collapse under different conditions \cite{CHB1,CHB2}, with particular interest in charged static BH solutions featuring flat horizons \cite{CHB3,CHB4}. A key distinction arises when comparing higher-dimensional Ricci-flat Reissner-Nordstr\"{o}m (RN) BHs to charged BTZ solutions, particularly in three dimensions. In the case of a three-dimensional RN spacetime with a flat horizon, the temporal component of the gauge potential, $A_t$, remains constant, leading to a vanishing electric field, whereas in BTZ BHs, $A_t$ follows a logarithmic dependence on the radial coordinate $r$, generating an electric field that diminishes as $r^{-1}$ \cite{CHB5,CHB6}. Furthermore, the charge contribution in the metric function of RN solutions scales as $r^{-2(n-2)}$ in higher dimensions ($\mathcal{D}$), but in three dimensions ($n = 2$), this term reduces to a constant, effectively eliminating the electric charge and resulting in an uncharged solution \cite{CHB7}. Also, BTZ BH models retain a logarithmic charge term in their metric function, ensuring a non-zero conserved electric charge even in three dimensions (3D) \cite{CHB8}. In this case, this fundamental difference illustrates that while higher-dimensional ($\mathcal{D}$) RN BHs lose their charge properties when reduced to three dimensions, BTZ BHs maintain a distinct electrically charged structure, emphasizing the unique behavior of these solutions in lower-dimensional ($\mathcal{D}$) gravity \cite{CHB9}. Moreover, recent breakthroughs in understanding quantum gravity corrections to BH thermodynamics through the Generalized Uncertainty Principle (GUP) have revealed that quantum effects can significantly suppress Hawking radiation and potentially lead to BH remnants \cite{isz09,isz10}. These quantum corrections provide crucial insights into the final stages of BH evaporation and the resolution of the information paradox.

Our study tests the dynamics of particles and waves in a three-dimensional charged BTZ-like BH model modified by the presence of disclinations, inspired by Bopp–Podolsky (BP) electrodynamics. Also, by analyzing both null and timelike geodesics, we reveal how key geometric parameters, such as the black hole mass, electric charge, cosmological constant, BP coupling, and disclination parameter, shape the effective potential and influence photon orbits, including conditions for the formation and stability of circular trajectories. In this case, we further investigate the propagation of massless scalar fields by reformulating the Klein-Gordon equation into a Schrödinger-like form, identifying an effective potential that governs wave behavior in the curved background. Extending our analysis to quantum gravity effects, we derive the generalized uncertainty principle (GUP)-corrected Hawking temperature, show how quantum corrections lead to a suppression of thermal radiation, and suggest the possibility of BH remnants. Finally, we compute the Keplerian frequency for circular orbits and the dynamic influence of parameters in the BH model.

The structure of this paper is as follows: In Sec. (\ref{sec2}), we introduce the BTZ-like BH solution with disclinations, and the geometric properties of spacetime are detailed in Sec. (\ref{sec3}). In Sec. (\ref{sec4}), we analyze the massless scalar field by studying the Klein-Gordon equation. Section (\ref{sec5}) is devoted to investigating the GUP-corrected Hawking temperature of the charged BTZ-like BH, as well as the Keplerian frequencies associated with these charged configurations (see Sec. (\ref{sec6})). Finally, we summarize and discuss our results in the conclusion section (\ref{sec7}).

\section{Charged BTZ-like black hole solution with disclinations}\label{sec2}

A circularly symmetric and static three-dimensional charged BH spacetime pierced by disclinations is described by the following line element \cite{EPJC} (see Eq. (26) therein):
\begin{equation}
    ds^2=-f(r)\,dt^2+\frac{dr^2}{g(r)}+\beta^2\,r^2\,d\phi^2,\label{bb1}
\end{equation}
where the metric functions $f(r)$ and $g(r)$ are given by:
\begin{align}
    f(r) & =-M-\Lambda\,r^2-\frac{4\,b^2\, Q^2\, M}{r^2}-2\,Q^2\, \ln\left( \frac{r}{r_0} \right),\nonumber\\
    g(r) &=-M-\Lambda\,r^2+8\,b^2\,Q^2\,\Lambda-\frac{4\,b^2\, Q^2\, M}{r^2}-2\,Q^2\, \ln\left( \frac{r}{r_0} \right).\label{bb2}
\end{align}

Here $M$ represents the BH mass, $b$ is the BP parameter, $Q$ denotes the electric charge, $\Lambda$ is the cosmological constant, $r_0$ is a positive scale parameter, and $\beta=(1-4\,\lambda)$ is the disclination parameter with $\lambda$ being the linear mass density of the cosmic strings \cite{CS1,CS2,CS3}. The disclination parameter $\beta$ is introduced by redefining the angular coordinate $\phi \to \phi' \equiv \beta\,\phi$, which effectively modifies the angular periodicity and creates a conical singularity at the origin. This geometric defect represents the presence of topological defects in the spacetime, similar to cosmic strings in cosmological models.

A crucial feature of this solution is that $f(r) \neq g(r)$ due to the nonminimal coupling of the BP electromagnetic field to gravity. This asymmetry between the metric functions is a distinctive characteristic that distinguishes BP electrodynamics from standard Maxwell theory. The BP parameter $b^2$ introduces additional nonlinear corrections that modify both the electromagnetic and gravitational sectors of the theory.

The BH horizons is located at the largest positive root of $g(r) = 0$, which approximately gives:
\begin{equation}
r_{ \pm}^2=\frac{Q^2 \pm \sqrt{Q^4+16 Q^2 b^2 M}}{2(-\Lambda)} \approx \frac{Q^2 \pm\left(Q^2+8 b^2 M\right)}{2(-\Lambda)},
\end{equation}
in which the event horizon is represented by $r_+=r_h$. These horizon radii demonstrate the interplay between charge, mass, cosmological constant, and the BP coupling parameter. When $b^2 \to 0$, we recover the standard result $r_{+} =\frac{Q^2}{(-\Lambda)}=Q^2\,\ell^2_\text{AdS}$, where $\ell_\text{AdS}=\sqrt{-\frac{1}{\Lambda}}$ represents the AdS radius.

\begin{longtable}{|>{$}c<{$}|>{$}c<{$}|c|c|}
\hline
\rowcolor{gray!50}
\mathbf{b} & \mathbf{Q} & \textbf{Horizon(s)} & \textbf{Classification} \\
\hline
\endfirsthead
\hline
\rowcolor{gray!50}
\mathbf{b} & \mathbf{Q} & \textbf{Horizon(s)} & \textbf{Classification} \\
\hline
\endhead

0.0 & 0.5 & $[0.13634539,\ 2.7426899]$ & Non-extremal BH \\
\hline
0.0 & 1.0 & $[0.63118207,\ 4.4689531]$ & Non-extremal BH \\
\hline
0.0 & 1.5 & $[0.82535105,\ 6.9798324]$ & Non-extremal BH \\
\hline
0.0 & 2.0 & $[0.90057285,\ 9.8164302]$ & Non-extremal BH \\
\hline
0.0 & 2.5 & $[0.93615153,\ 12.824117]$ & Non-extremal BH \\
\hline
0.0 & 3.0 & $[0.95560649,\ 15.944240]$ & Non-extremal BH \\
\hline
0.01 & 0.5 & $[0.13482029,\ 2.7427482]$ & Non-extremal BH \\
\hline
0.01 & 1.0 & $[0.63078261,\ 4.4690874]$ & Non-extremal BH \\
\hline
0.01 & 1.5 & $[0.82502269,\ 6.9800087]$ & Non-extremal BH \\
\hline
0.01 & 2.0 & $[0.90026620,\ 9.8166412]$ & Non-extremal BH \\
\hline
0.01 & 2.5 & $[0.93585458,\ 12.824361]$ & Non-extremal BH \\
\hline
0.01 & 3.0 & $[0.95531475,\ 15.944517]$ & Non-extremal BH \\
\hline
0.1 & 0.5 & $[2.7485010]$ & Extremal or Single Root BH \\
\hline
0.1 & 1.0 & $[0.10781792,\ 0.58782435,\ 4.4823452]$ & Non-extremal BH \\
\hline
0.1 & 1.5 & $[0.097676086,\ 0.79107004,\ 6.9974189]$ & Non-extremal BH \\
\hline
0.1 & 2.0 & $[0.094860947,\ 0.86881534,\ 9.8374952]$ & Non-extremal BH \\
\hline
0.1 & 2.5 & $[0.093652280,\ 0.90549468,\ 12.848515]$ & Non-extremal BH \\
\hline
0.1 & 3.0 & $[0.093018215,\ 0.92553303,\ 15.971910]$ & Non-extremal BH \\
\hline
0.5 & 0.5 & $[2.8801561]$ & Extremal or Single Root BH \\
\hline
0.5 & 1.0 & $[4.7819624]$ & Extremal or Single Root BH \\
\hline
0.5 & 1.5 & $[7.3982004]$ & Extremal or Single Root BH \\
\hline
0.5 & 2.0 & $[10.322900]$ & Extremal or Single Root BH \\
\hline
0.5 & 2.5 & $[13.414314]$ & Extremal or Single Root BH \\
\hline
0.5 & 3.0 & $[16.616161]$ & Extremal or Single Root BH \\
\hline
\caption{\footnotesize Horizon radii for different values of the BP parameter $b$ and BH charge $Q$, illustrating the nature and multiplicity of the BH roots. The other physical parameters are chosen as $M=r_0=1$ and $\Lambda=-0.2$.}
\label{tab:horizons}
\end{longtable}

Table \ref{tab:horizons} reveals the profound influence of the BP parameter $b$ on the horizon structure of charged BTZ-like BHs with disclinations. For the standard case ($b = 0$), all charge configurations exhibit non-extremal black holes with two distinct horizons, representing the characteristic inner and outer horizons found in charged BH solutions. When the BP parameter increases to $b = 0.1$, the horizon structure becomes more complex, with most configurations still maintaining non-extremal behavior but showing additional root structures, except for the case $Q = 0.5$ which exhibits extremal or single-root behavior. Remarkably, at the higher BP coupling $b = 0.5$, all charge configurations transition to extremal or single-root BHs, indicating that strong nonlinear electromagnetic effects fundamentally alter the causal structure by eliminating the distinction between inner and outer horizons. This transition suggests that the BP corrections can drive the BH toward extremality, potentially affecting its thermodynamic stability and evaporation properties.

The electromagnetic field tensor in BP electrodynamics is given by:
\begin{equation}
F_{\mu \nu}=E(r)\left[\delta_\mu^1 \delta_\nu^0-\delta_\mu^0 \delta_\nu^1\right],
\end{equation}
where $E(r)$ represents the two-dimensional spatial electric field. The electric field component has been found by Ref. \cite{EPJC} as follows 
\begin{equation}
E(r)=\frac{Q}{r}\left(1+2 b^2 \Lambda\right)+\frac{4 b^2 Q^3}{r^3}.
\end{equation}
This expression demonstrates the modification of the standard Coulomb-like electric field due to the BP nonlinear electromagnetic corrections, which introduce both the cosmological constant dependence and higher-order charge terms that are characteristic of 3D charged BH solutions in BP electrodynamics. The energy-momentum tensor for the BP electromagnetic field takes the form:
\begin{equation}
    T_{\mu\nu} = \frac{1}{4\pi}\left[F_{\mu\alpha}F_\nu^{\,\alpha} - \frac{1}{4}g_{\mu\nu}F_{\alpha\beta}F^{\alpha\beta}\right] + T_{\mu\nu}^{(BP)},
\end{equation}
where $T_{\mu\nu}^{(BP)}$ is given by 
\begin{equation}
T_{\mu \nu}^{(BP)}=\frac{b^2}{2 \pi}\left[-\frac{1}{4} g_{\mu \nu} \nabla^\beta F^{\alpha \gamma} \nabla_\beta F_{\alpha \gamma}+F_{(\mu}^\gamma \nabla^\beta \nabla_\beta F_{\nu) \gamma}+F_{\gamma(\mu} \nabla_\beta \nabla_{\nu)} F^{\beta \gamma}-\nabla_\beta\left(F_\gamma{ }^\beta \nabla_{(\mu} F_{\nu)}{ }^\gamma\right)\right],
\end{equation}

which arises from the nonminimal coupling seen in the Lagrangian density of BP electrodynamics \cite{EPJC}. These corrections are responsible for the difference between $f(r)$ and $g(r)$ in the metric functions. The causal structure of this spacetime is determined by the location of the horizons and the behavior of null geodesics \cite{isadd3,isadd4}. 

In the limit $b^2 \to 0$, one can recover the standard charged BTZ solution with disclinations given by the following line element:
\begin{equation}
    ds^2=-A(r)\,dt^2+\frac{dr^2}{A(r)}+\beta^2\,r^2\,d\phi^2,\label{bb3}
\end{equation}
where the metric function $A(r)$ is now given by (setting $r_0=1$):
\begin{align}
    A(r)=-M-\Lambda\,r^2-2\,Q^2\, \ln r.\label{bb4}
\end{align}

The thermodynamic properties of this BH solution are governed by the surface gravity $\kappa$ at the horizon, which determines the Hawking temperature through the relation $T_H = \kappa/(2\pi)$. For this specific charged BTZ-like BH with disclinations, the Hawking temperature takes the explicit form \cite{wald}:
\begin{align}
T_H&=\frac{1}{4 \pi}  \frac{f^{\prime}\left(r_h\right)  \sqrt{g\left(r_h\right)}}{\sqrt{f\left(r_h\right)}}, \notag \\ 
&=\frac{\left(-2 \Lambda r_{-} h+\frac{8 b^2 Q^2 M}{r_{-} h^3}-\frac{2 Q^2}{r_{-} h}\right) \sqrt{-M-\Lambda r_{-} h^2+8 b^2 Q^2 \Lambda-\frac{4 b^2 Q^2 M}{r_{-} h^2}-2 Q^2 \ln \left(\frac{r_{-} h}{r_0}\right)}}{4 \sqrt{-M-\Lambda r_{-} h^2-\frac{4 b^2 Q^2 M}{r_{-} h^2}-2 Q^2 \ln \left(\frac{r_{-} h}{r_0}\right)} \pi}, \label{isTH} 
\end{align}
where $r_H$ denotes the event horizon and the prime symbol denotes the derivative with respect to $r$. This expression incorporates the distinctive features of the BP electrodynamics through the metric functions $f(r)$ and $g(r)$, reflecting the non-trivial coupling between the electromagnetic field and gravity in this modified theory. The presence of the disclination parameter $\beta$ and the BP coupling $b^2$ significantly modifies these thermodynamic quantities compared to the standard BTZ BH.

\section{Geometric Properties: Null geodesics}\label{sec3}

In this section, we examine light propagation in the geometric background described by the metric given in \eqref{bb2}. We begin by studying the geodesic equation to determine the light ray trajectories, which provides fundamental insights into the causal structure and optical properties of our charged BTZ-like BH with disclinations. The analysis of null geodesics is particularly important for understanding observational signatures such as photon rings, shadows, and gravitational lensing effects that could be detected by advanced instruments like the EHT \cite{Akiyama2019a,Akiyama2019b,iszx01,iszx02}.

We take the Lagrangian density function as:
\begin{equation}
\mathcal{L} = g_{\mu \nu}\,\dot{x}^{\mu}\,\dot{x}^{\nu},\label{cc1} 
\end{equation}
where the dot represents the ordinary derivative with respect to the affine parameter $\tau$ of the curve. Geodesics are obtained as solutions of the Euler-Lagrange equation:
\begin{equation}
\frac{\partial \mathcal{L}}{\partial x^{\mu}} - \frac{d}{d\tau} \left( \frac{\partial \mathcal{L}}{\partial \dot{x}^{\mu}} \right) = 0.\label{cc2}
\end{equation}

The Lagrangian density function using the metric (\ref{bb1}) becomes:
\begin{eqnarray}
\mathcal{L} = -f(r)\,\dot{t}^2+\frac{\dot{r}^2}{g(r)}+\beta^2\,r^2\,\dot{\phi}^2,\label{cc3}
\end{eqnarray}

From the above function, we observe that $\mathcal{L}$ does not explicitly depend on $t$ and $\phi$. Hence, there are two conserved quantities associated with these cyclic coordinates:
\begin{equation}
\mathrm{E} =f(r)\,\dot{t}\quad, \quad \mathrm{L}=\frac{p_{\phi}}{\beta}=\beta\,r^2\, \dot{\phi},\label{cc5}
\end{equation}
where $\mathrm{E}$ is the conserved energy and $\mathrm{L}_z=\mathrm{L}$ is the conserved angular momentum about the symmetry axis.

With these conserved quantities, we can rewrite Eq. (\ref{cc3}) in the form of an effective one-dimensional motion:
\begin{equation}
    k(r)\,\left(\frac{dr}{d\tau}\right)^2+V_\text{eff}(r)=\mathrm{E}^2,\quad\quad k(r)=\frac{f(r)}{g(r)}\label{cc6}
\end{equation}
which is analogous to the one-dimensional equation of motion of a particle of unit mass possessing energy $\mathrm{E}^2$ and moving in an effective potential $V_\text{eff}(r)$. This potential is given by:
\begin{align}
    V_\text{eff}(r)=f(r)\,\left[-\varepsilon+\frac{\mathrm{L}^2}{r^2}\right]=\left[-M-\Lambda\,r^2-\frac{4\,b^2\, Q^2\, M}{r^2}-2\,Q^2\, \ln\left( \frac{r}{r_0} \right)\right]\,\left(-\varepsilon+\frac{\mathrm{L}^2}{r^2}\right),\label{cc7}
\end{align}
where $\varepsilon=0$ for null geodesics and $\varepsilon=-1$ for timelike geodesics.

\begin{figure}[ht!]
    \centering
    \includegraphics[width=0.45\linewidth]{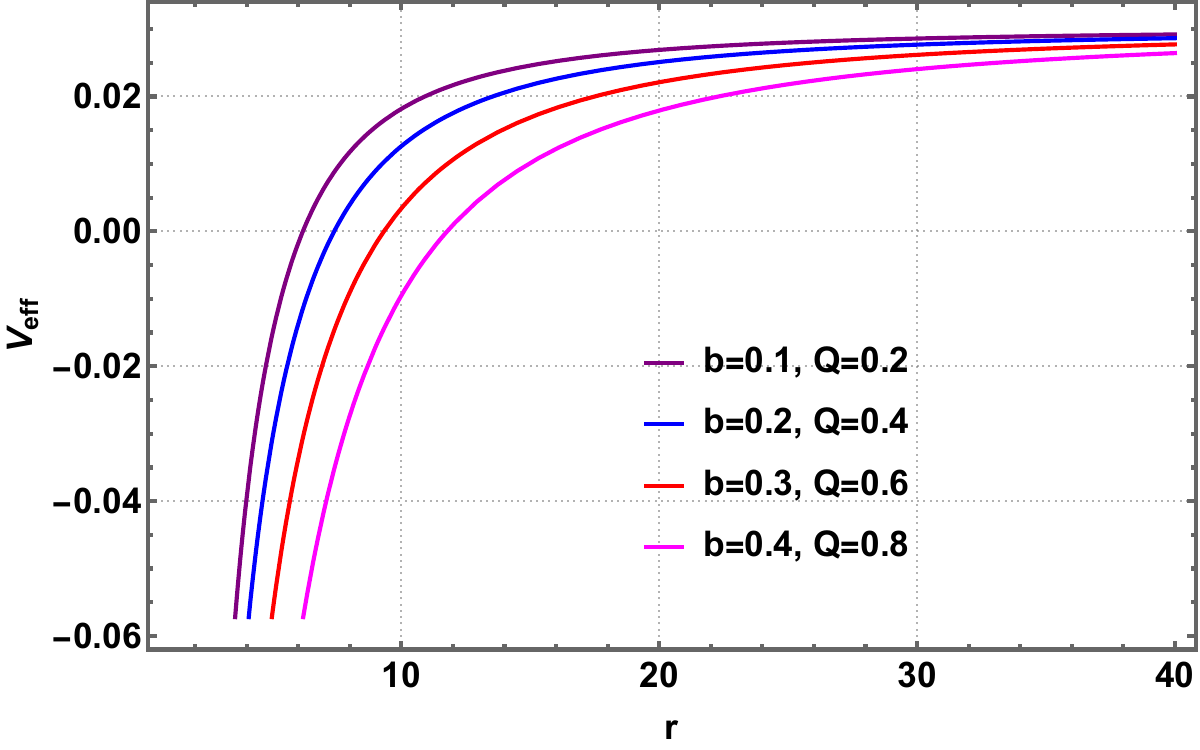}
    \caption{Illustration of the effective potential for null geodesics as a function of $r$, with simultaneous variation of the electric charge $Q$ and the non-minimal coupling parameter $b$. Here, we set the BH mass $M=1$, the cosmological constant $\Lambda=-0.3$, the scale parameter $r_0=1$, and the angular momentum $\mathrm{L}=1$ fixed, all in natural units. While these values are held constant here, varying $M, r_0$ and $\mathrm{L}$ would further show their influence on the effective potential.}
    \label{fig:potential-1}
    \hfill\\
    \includegraphics[width=0.45\linewidth]{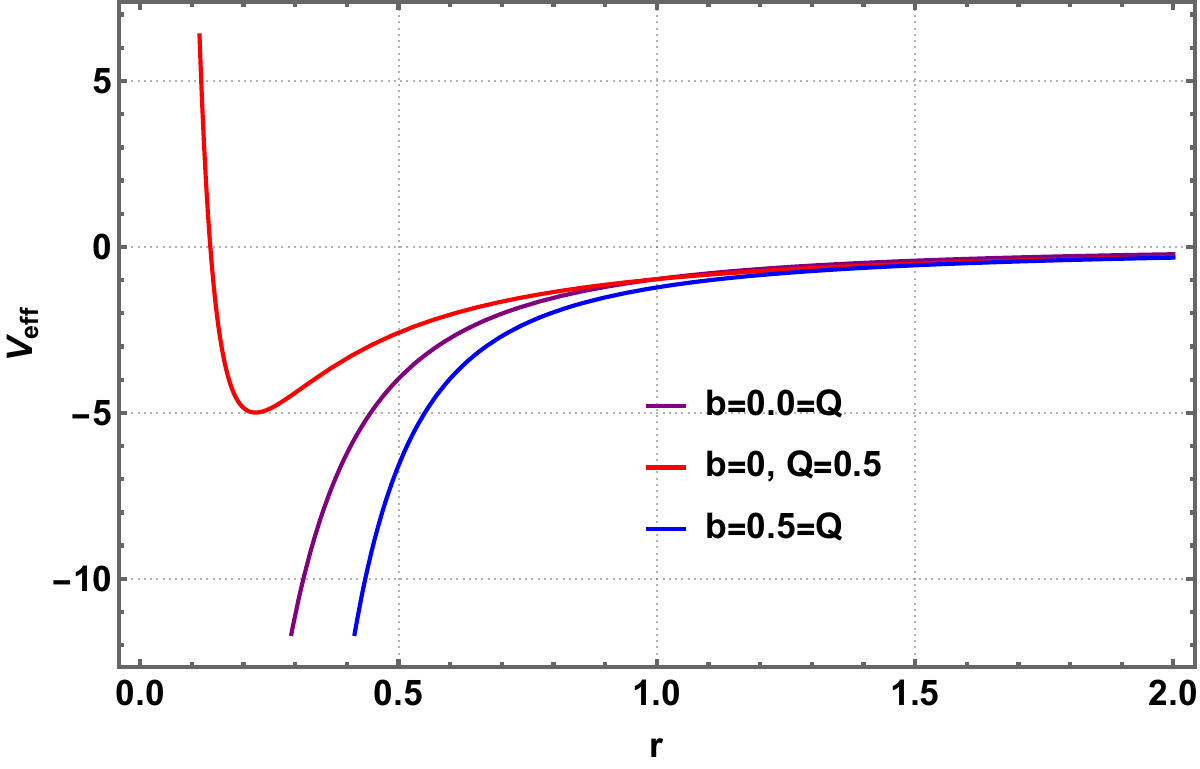}
    \caption{Comparison of the effective potential $V_{\text{eff}}(r)$ for null geodesics under three different scenarios: (i) uncharged BTZ BH without coupling (purple curve), (ii) charged BTZ BH without coupling (red curve), and (iii) charged BTZ BH with non-minimal coupling (blue curve). The parameters are fixed as the BH mass $M = 1$, the cosmological constant $\Lambda = -0.3$, $r_0 = 1$, the scale parameter $r_0=1$, and the angular momentum $\mathrm{L} = 1$, all in natural units. This comparison highlights the distinct influence of electric charge and non-minimal coupling on the effective potential profile, particularly in modifying the potential barrier's height and shape, which in turn affects the nature of photon orbits.}
    \label{fig:potential-2}
\end{figure}

From the above expression, it becomes evident that the effective potential governing the geodesic motion of test particles is influenced by several key geometric parameters. These include the BH mass $M$, electric charge $Q$, cosmological constant $\Lambda$, disclination parameter $\beta$, and the BP coupling parameter $b^2$, while keeping the scale parameter $r_0$ fixed. The interplay between these parameters creates a rich phenomenology in the orbital dynamics, as demonstrated in the accompanying figures.

\subsection{Photon and Particle Trajectories}

Now, we study trajectories of photon and timelike particles in the curved spacetime and analyze how various parameters influence the geodesic motion. The equation of photon orbits using Eqs. (\ref{cc5}) and (\ref{cc6}) is given by:
\begin{equation}
    \left(\frac{dr}{d\phi}\right)^2=\beta^2\,r^4\,\left(1+\frac{8\,b^2\,Q^2\,\Lambda}{f(r)}\right)\,\left(\frac{1}{\gamma^2}-\frac{f(r)}{r^2}\right),\label{ee2}
\end{equation}
where $\gamma=\mathrm{L}/\mathrm{E}$ is the impact parameter for the photon ray. This expression clearly shows how the BP corrections, encoded in the term $8\,b^2\,Q^2\,\Lambda/f(r)$, modify the standard geodesic equations and influence the photon trajectories \cite{iszx03,iszx04}.

In the limit $b^2 \to 0$, the photon trajectory equation reduces to the standard charged BTZ form:
\begin{equation}
    \left(\frac{dr}{d\phi}\right)^2=\beta^2\,r^4\,\left(\frac{1}{\gamma^2}-\frac{A(r)}{r^2}\right).\label{ee2b}
\end{equation}

Similarly, the equation for timelike particle trajectories is given by:
\begin{eqnarray}
    \left(\frac{dr}{d\phi}\right)^2=\beta^2\,r^4\,\left(1+\frac{8\,b^2\,Q^2\,\Lambda}{f(r)}\right)\,\left[\frac{\mathrm{E}^2}{\mathrm{L}^2}-\left(\frac{1}{\mathrm{L}^2}+\frac{1}{r^2}\right)\,f(r)\right],\label{ee3}
\end{eqnarray}
where $\mathrm{L}$ represents the specific angular momentum and $\mathrm{E}$ represents the specific energy of timelike particles.

In the limit $b^2 \to 0$, the particle trajectory reduces to:
\begin{eqnarray}
    \left(\frac{dr}{d\phi}\right)^2=\beta^2\,r^4\,\left[\frac{\mathrm{E}^2}{\mathrm{L}^2}-\left(\frac{1}{\mathrm{L}^2}+\frac{1}{r^2}\right)\,A(r)\right].\label{ee3b}
\end{eqnarray}

\subsection{Circular Photon Orbits and Critical Impact Parameter}

Now, we study circular photon orbits analogous to the analysis done {\color{red} in \cite{EPJC2}} using the effective potential for null geodesics. From Eq. (\ref{cc7}), the effective potential for null geodesics is:
\begin{equation}
    V_\text{eff}(r)=f(r)\,\frac{\mathrm{L}^2}{r^2},\label{ee4}
\end{equation}

For circular photon orbits of radius $r=\text{const}$, we must have the conditions $\dot{r}=0$ and $\ddot{r}=0$. The first condition implies:
\begin{equation}
    \mathrm{E}^2=\frac{\mathrm{L}^2}{r^2}\,f(r)\label{ee5}
\end{equation}
which gives us the critical impact parameter for photon rays. This critical impact parameter is:
\begin{equation}
    \gamma_c=\frac{r}{\sqrt{f(r)}}=\frac{r}{\sqrt{-M-\Lambda\,r^2-\frac{4\,b^2\, Q^2\, M}{r^2}-2\,Q^2\, \ln\left( \frac{r}{r_0} \right)}}.\label{ee6}
\end{equation}

Depending upon the relation between the impact parameter $\gamma=\mathrm{L}/\mathrm{E}$ and the critical impact parameter $\gamma_c$, the photon rays are either captured by the BH or escape to infinity. In the special case $\gamma=\gamma_c$, photon particles move along circular null orbits, which are crucial for understanding the BH shadow and photon ring formation \cite{iszx05,iszx06}.

From the above expression, it becomes evident that the critical impact parameter is influenced by key geometric parameters, including the BH mass $M$, electric charge $Q$, cosmological constant $\Lambda$, and the BP coupling parameter $b^2$, while keeping the scale parameter $r_0$ fixed.

Moreover, the second condition $\ddot{r}=0$ implies the following relation:
\begin{align}
    V'_\text{eff}(r)=0\Rightarrow r\,f'-2\,f=0\Rightarrow M + \frac{8\,b^2\, Q^2\, M}{r^2} -Q^2 + 2\,Q^2\, \ln\left( \frac{r}{r_0} \right)=0.\label{ee7}
\end{align}

From this relation, it becomes evident that the radius $r=r_\text{cpo}$ of the circular photon orbits (CPOs) is influenced primarily by the BH mass $M$, electric charge $Q$, and the BP coupling parameter $b^2$, while keeping the scale parameter $r_0$ fixed. However, in the limit $b^2 \to 0$, we find this radius $r_\text{cpo}=r_0\,\exp\left(\frac{1}{2}-\frac{M}{2\,Q^2}\right)$.

It is very difficult to solve analytically the above equation (\ref{ee7}) due to the logarithmic function. However, we can determine numerically the radius of the circular photon orbits by setting suitable values of other parameters. Table~\ref{tab:r-vs-Q-b} presents the numerical values of this radius for various choices of the charge $Q$ and the BP coupling parameter $b$. The results reveal a complex interplay between the electric charge and BP coupling effects on the circular photon orbit radius. For a fixed BP parameter $b = 0.1$, increasing the charge $Q$ from $0.8$ to $1.6$ leads to a monotonic increase in $r_\text{cpo}$, indicating that higher charge values push the photon orbits further from the BH center. Conversely, for a larger BP coupling $b = 0.2$, the behavior becomes more intricate, with the orbit radius initially decreasing from $Q = 0.8$ to $Q = 1.0$ before subsequently increasing for higher charge values. This non-monotonic behavior demonstrates that the BP corrections introduce competing effects that can either enhance or suppress the influence of electric charge on photon trajectories, depending on the relative magnitudes of the parameters. The overall trend suggests that both charge and BP coupling work together to modify the effective gravitational potential experienced by photons, with implications for observable phenomena such as photon ring formation and gravitational lensing signatures.

\begin{table}[h!]
\centering
\begin{tabular}{|c|c|c|}
\hline
$Q \backslash b$ & $0.1$ & $0.2$ \\
\hline
$0.8$ & 0.160850 & 0.565686 \\
$1.0$ & 0.957290 & 0.443781 \\
$1.2$ & 1.129077 & 0.989382 \\
$1.4$ & 1.244943 & 1.126047 \\
$1.6$ & 1.325682 & 1.217419 \\
\hline
\end{tabular}
\caption{Numerical solutions for $r_\text{cpo}$ for various values of $Q$ and $b$, based on the equation $M + \frac{8\, b^2\, Q^2\, M}{r^2} - Q^2 + 2\, Q^2\,\ln\left( \frac{r}{r_0} \right) = 0$ with $M=1$, $r_0=1$.}
\label{tab:r-vs-Q-b}
\end{table}

Figure \ref{fig:photon} provides a clear visualization of the influence of both the BP coupling and charge parameters on the circular photon orbit.

\begin{figure}[ht!]
    \centering   {\centering{}\includegraphics[width=0.46\linewidth]{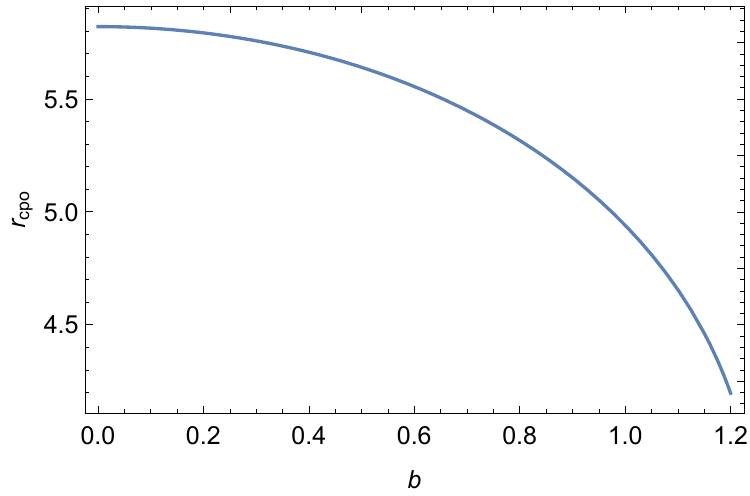}}\quad\quad
{\centering{}\includegraphics[width=0.45\linewidth]{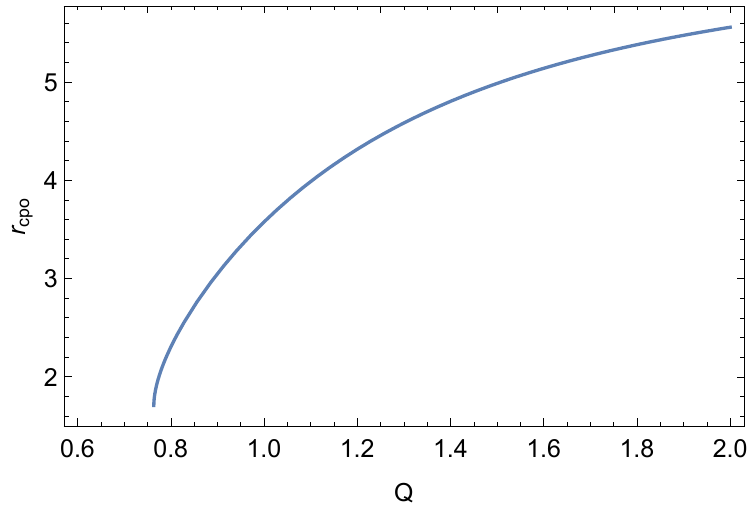}}\\
    \caption{Plot of circular photon orbit $r_{cpo}$ for different values of BP coupling parameter $b$ with fixed $Q=2$ (left panel) and for different values of charge parameter $Q$ with fixed $b=0.6$ (right panel). Here, we set $M=1$ and $r_0=1$.}
    \label{fig:photon}
\end{figure}

\begin{figure}[ht!]
    \centering   {\centering{}\includegraphics[width=0.46\linewidth]{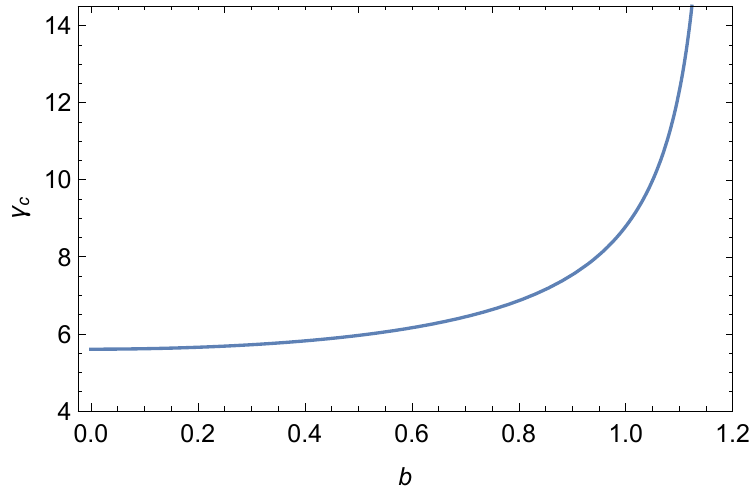}}\quad\quad
{\centering{}\includegraphics[width=0.45\linewidth]{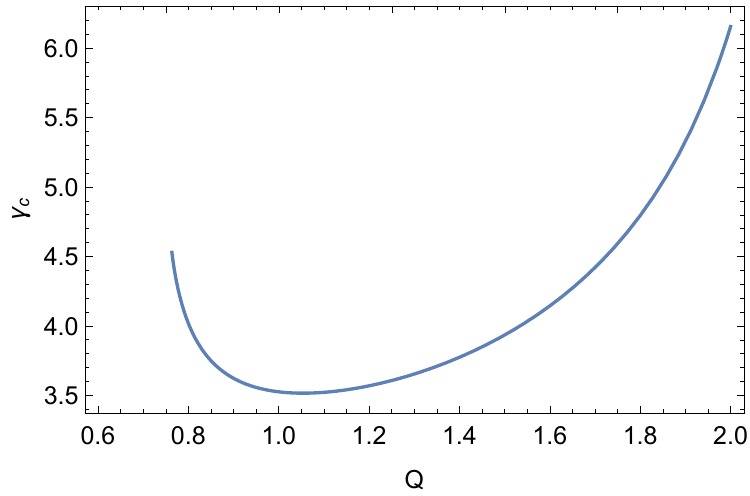}}\\
    \caption{Plot of impact parameter $\gamma_c$ for different values of BP coupling parameter $b$ with fixed $Q=2$ (left panel) and for different values of charge parameter $Q$ with fixed $b=0.6$ (right panel). Here, we set $M=1$ and $r_0=1$.}
    \label{fig:impact}
\end{figure}

\subsection{Stability Analysis and Photon Forces}

To study the stability of the circular photon orbits, we need to examine whether $V''_\text{eff}(r)$ is positive or negative. This is given by:
\begin{equation}
    V''_\text{eff}(r)=\mathrm{L}^2\,\left(\frac{f''(r)}{r^2}-\frac{2\,f(r)}{r^4}\right)=\mathrm{L}^2\,\left[ 
- \frac{4\,\Lambda}{r^2} 
- \frac{32\,b^2\, Q^2\, M}{r^6} 
+ \frac{2\, Q^2 - 2\,M - 4\, Q^2\, \ln\left( \frac{r}{r_0} \right)}{r^4}\right].\label{condition}
\end{equation}
At the radius $r=r_\text{cpo}$, if $V''_\text{eff}(r)<0$, then the circular orbits are stable; otherwise, these orbits are unstable. This stability criterion is crucial for understanding the long-term behavior of photon trajectories and the formation of observable features such as photon rings \cite{iszx07,iszx08}.

The force on photon particles in the given gravitational field is defined in terms of the effective potential for null geodesics $V_\text{eff}(r)$. The expression is:
\begin{equation}
    \mathrm{F}_\text{ph}=-\frac{1}{2}\,\frac{dV_\text{eff}}{dr}=\frac{\mathrm{L}^2}{r^3}\,\left[-M - \frac{8\, b^2\, Q^2\, M}{r^2} + Q^2 - 2\,Q^2\, \ln\left( \frac{r}{r_0} \right)\right].\label{force}
\end{equation}

\begin{figure}[ht!]
    \centering
    \includegraphics[width=0.46\linewidth]{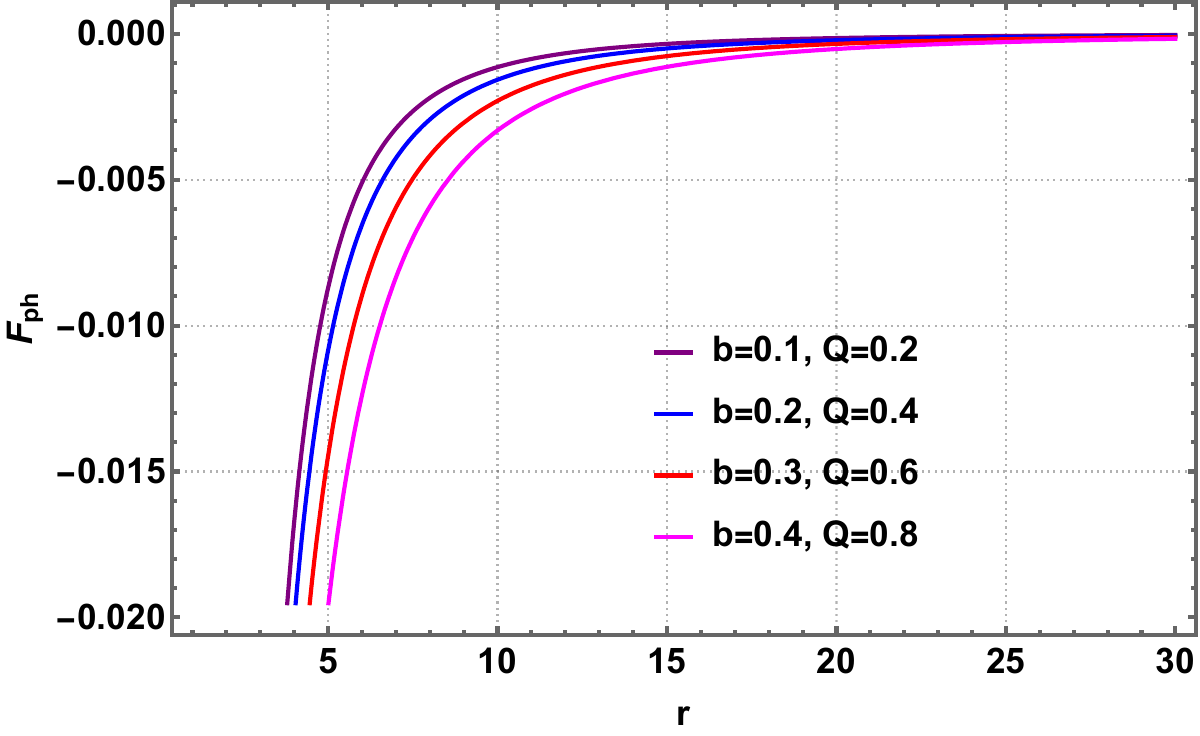}
    \caption{Illustration of the force on photon particles as a function of $r$, with simultaneous variation of the electric charge $Q$ and the BP coupling parameter $b$. Here, we set the BH mass $M=1$, the cosmological constant $\Lambda=-0.3$, the scale parameter $r_0=1$, and the angular momentum $\mathrm{L}=1$ fixed, all in natural units. While these values are held constant here, varying $M, r_0$ and $\mathrm{L}$ would further show their influence on the nature of the force.}
    \label{fig:force-1}
    \hfill\\
    \includegraphics[width=0.45\linewidth]{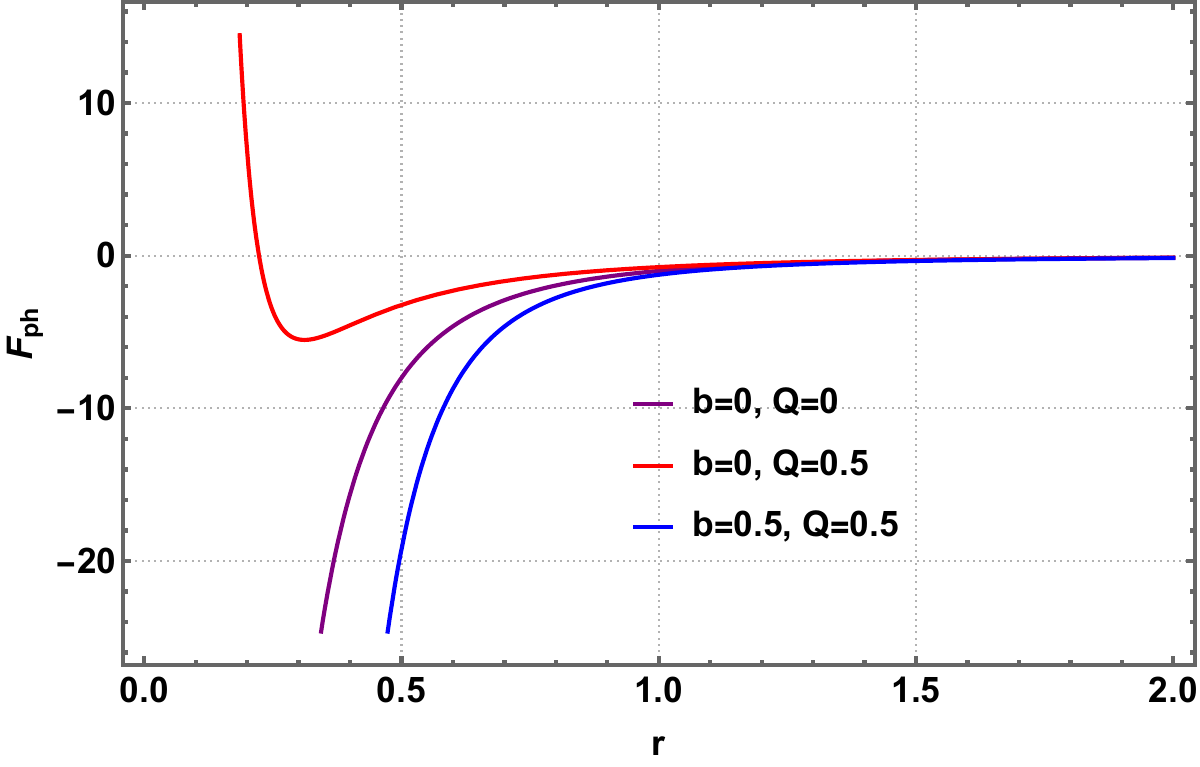}
    \caption{Comparison of the force $F_{\text{ph}}(r)$ on the photon particles under three different scenarios: (i) uncharged BTZ BH without coupling (purple curve), (ii) charged BTZ BH without coupling (red curve), and (iii) charged BTZ BH with BP coupling (blue curve). The parameters are fixed as the BH mass $M = 1$, the cosmological constant $\Lambda = -0.3$, $r_0 = 1$, the scale parameter $r_0=1$, and the angular momentum $\mathrm{L} = 1$, all in natural units. This comparison highlights the distinct influence of electric charge and BP coupling on the dynamics of photon particles, particularly their attractive or repulsive nature, which determines whether they are captured or escape from the BH region.}
    \label{fig:force-2}
\end{figure}

From the above expression, it becomes evident that the force on photon particles is influenced by the key geometric parameters, including the BH mass $M$, electric charge $Q$, and the BP coupling parameter $b^2$, as well as the scale parameter $r_0$. Moreover, the angular momentum $\mathrm{L}$ significantly alters the force on the photon particles.

\subsection{Angular Velocity and Black Hole Shadow}

For circular null geodesics, we find the angular velocity given by:
\begin{equation}
    \omega=\frac{\dot{\phi}}{\dot{t}}\Big{|}_{r=r_\text{cpo}}=\frac{1}{\beta}\,\frac{\sqrt{-M-\Lambda\,r^2-\frac{4\,b^2\, Q^2\, M}{r^2}-2\,Q^2\, \ln\left( \frac{r}{r_0} \right)}}{r}\Bigg{|}_{r=r_\text{cpo}}.\label{ee11}
\end{equation}

From this expression, it becomes evident that the angular speed is influenced by the key geometric parameters, including the BH mass $M$, electric charge $Q$, and the BP coupling parameter $b^2$, as well as the scale parameter $r_0$. Moreover, the disclination parameter $\beta$ directly modifies this speed on the circular orbits, reflecting the topological nature of the spacetime defect.

Finally, we focus on the BH shadow and examine how various parameters alter the size of the shadow. This shadow radius can be determined using the following relation \cite{iszx09,iszx10}:
\begin{equation}
    R_\text{s}=\frac{r_\text{cpo}}{\sqrt{f(r_\text{cpo})}}=\frac{r_\text{cpo}}{\sqrt{-M-\Lambda\,r^2_\text{cpo}-\frac{4\,b^2\, Q^2\, M}{r^2_\text{cpo}}-2\,Q^2\, \ln\left( \frac{r_\text{cpo}}{r_0} \right)}}.\label{radius-1}
\end{equation}

\begin{table}[h!]
\centering
\begin{minipage}{0.5\textwidth}
\centering
\begin{tabular}{|c|c|c|c|}
\hline
$r_\text{cpo}$ & $b$ & $Q$ & $R_\text{s}$ \\
\hline
3.5 & 0.2 & 0.5 & 2.4473 \\
3.5 & 0.4 & 0.5 & 2.4532 \\
3.5 & 0.6 & 0.5 & 2.4631 \\
3.5 & 0.8 & 0.5 & 2.4771 \\
3.5 & 1.0 & 0.5 & 2.4956 \\
3.5 & 1.2 & 0.5 & 2.5187 \\
3.5 & 1.4 & 0.5 & 2.5468 \\
3.5 & 1.6 & 0.5 & 2.5805 \\
3.5 & 1.8 & 0.5 & 2.6203 \\
\hline 
\end{tabular}
\caption{Shadow radius for varying $b$ (fixed $Q = 0.5$).} \label{istab3}
\end{minipage}
\hfill
\begin{minipage}{0.48\textwidth}
\centering
\begin{tabular}{|c|c|c|c|}
\hline
$r_\text{cpo}$ & $b$ & $Q$ & $R_\text{s}$ \\
\hline
3.5 & 0.1 & 0.1 & 2.15007 \\
3.5 & 0.1 & 0.2 & 2.18127 \\
3.5 & 0.1 & 0.3 & 2.23643 \\
3.5 & 0.1 & 0.4 & 2.32120 \\
3.5 & 0.1 & 0.5 & 2.44582 \\
3.5 & 0.1 & 0.6 & 2.62940 \\
3.5 & 0.1 & 0.7 & 2.91092 \\
3.5 & 0.1 & 0.8 & 3.38457 \\
3.5 & 0.1 & 0.9 & 4.36519 \\
\hline 
\end{tabular}
\caption{Shadow radius for varying $Q$ (fixed $b = 0.1$).} \label{istab4}
\end{minipage}
\end{table}

The shadow radii presented in Tables \ref{istab3} and \ref{istab4} reveal the significant influence of BP electrodynamics parameters on observable BH signatures. For fixed charge $Q = 0.5$, increasing the BP coupling parameter $b$ from $0.2$ to $1.8$ leads to a monotonic growth in the shadow radius from $2.4473$ to $2.6203$, demonstrating that stronger nonlinear electromagnetic effects systematically enlarge the apparent size of the BH shadow. Conversely, for fixed BP coupling $b = 0.1$, the shadow radius exhibits a more dramatic dependence on the electric charge, increasing from $2.15007$ at $Q = 0.1$ to $4.36519$ at $Q = 0.9$, indicating that the charge parameter has a more pronounced effect on the shadow size than the BP coupling. This pronounced charge dependence suggests that highly charged BTZ-like BHs would produce significantly larger and more easily detectable shadows in observational campaigns \cite{iszx11,iszx12}.

\section{Massless Scalar Field: The Klein-Gordon Equation} \label{sec4}

Scalar perturbations via the massless Klein-Gordon equation play a pivotal role in the study of BH physics, serving as a foundational tool for probing the stability of BH spacetimes under linear perturbations. These investigations offer crucial insights into the response of the background geometry to external disturbances and provide a window into the fundamental properties of curved spacetime \cite{iszx13,iszx14}. In particular, scalar field dynamics are instrumental in examining the quasinormal modes (QNMs) of BHs, which characterize the ringdown phase of gravitational wave emission and encode essential information about the geometry and horizon structure of the spacetime \cite{Kokkotas1999,Berti2009}.

The study of scalar perturbations becomes particularly compelling in the context of our charged BTZ-like BH with disclinations, where the interplay between BP electrodynamics, topological defects, and the AdS background creates a rich phenomenological landscape. Moreover, scalar perturbations are widely used to investigate various physical phenomena such as greybody factors, Hawking radiation, and superradiance, especially in higher-dimensional and asymptotically AdS BHs \cite{iszx15}. They also provide a simpler yet effective model to understand more complex field dynamics, such as electromagnetic or gravitational perturbations, without dealing with the full tensorial complexity inherent in these systems \cite{iszx16,iszx17}.

In many cases, the behavior of a test scalar field can reveal the presence of instabilities or novel physical effects arising from modifications to general relativity or the inclusion of exotic matter fields such as dark energy or quintessence \cite{Fernando2004,Cardoso2004}. For our specific geometry, the scalar field analysis will illuminate how the BP corrections and disclination parameters influence wave propagation and potentially affect the stability properties of the BH solution.

The massless Klein-Gordon equation in curved spacetime is described by the covariant wave equation:
\begin{equation}
   \frac{1}{\sqrt{-g}}\,\partial_{\mu}\left(\sqrt{-g}\,g^{\mu\nu}\,\partial_{\nu}\right)\,\Phi=0,\label{ff1}
\end{equation}
where $g=\det(g_{\mu\nu})$ is the determinant of the metric tensor, $g^{\mu\nu}$ is the inverse metric, and $\Phi$ is the massless scalar field. This equation represents the natural generalization of the flat-space wave equation to curved backgrounds and captures the essential physics of scalar wave propagation in strong gravitational fields \cite{iszx18,iszx19}.

Using the metric (\ref{bb1}), we arrive at the explicit form:
\begin{equation}
   -\frac{\partial^2 \Phi}{\partial t^2}+\frac{\sqrt{g(r)\,f(r)}}{r}\,\frac{\partial}{\partial r}\,\left(r\,\sqrt{f(r)\,g(r)}\,\frac{\partial}{\partial r}\right)\,\Phi+\frac{f(r)}{\beta^2\,r^2}\,\frac{\partial^2 \Phi}{\partial \phi^2}=0.\label{ff2}
\end{equation}

This equation explicitly demonstrates how the asymmetry between $f(r)$ and $g(r)$, arising from the BP electromagnetic corrections, manifests in the wave dynamics. The disclination parameter $\beta$ appears in the angular term, reflecting the modified angular structure of the spacetime due to the topological defect.

Since the given spacetime possesses cylindrical symmetry, we employ the following scalar field ansatz that exploits the symmetries of the problem:
\begin{equation}
   \Phi(t, r, \phi)=\exp(i\,\ell\,\phi)\,\exp(i\,\omega\,t)\,\psi(r),\label{ff3}
\end{equation}
where $\ell$ is the multipole number characterizing the angular dependence and $\omega$ is the frequency of the oscillation. This separation ansatz is particularly well-suited for axially symmetric spacetimes and has been extensively employed in studies of scalar perturbations around various BH geometries \cite{iszx20,iszx21}.

Substituting the scalar field ansatz into Eq. (\ref{ff2}) and performing the necessary algebraic manipulations, we obtain the radial differential equation:
\begin{align}
   \psi''(r)
+ \left[ \frac{1}{r} + \frac{1}{2} \left( \frac{f'(r)}{f(r)} + \frac{g'(r)}{g(r)} \right) \right] \psi'(r)
+ \left[ \frac{\omega^2}{f(r)} - \frac{\ell^2}{\beta^2 r^2} \right] \psi(r) = 0.\label{ff4}
\end{align}

Let us consider the following transformation
\begin{equation}
    \psi(r) = \frac{R(r)}{\left( r \sqrt{f(r)\,g(r)} \right)^{1/2}}
\end{equation}
This transformation eliminates the first-derivative term and reduces the equation to the canonical Schrödinger form \cite{iszx24,iszx25}. The resulting equation (\ref{ff4}) can be written as:
\begin{equation}
   R''(r)+\left(\omega^2-V_\text{eff}\right)\,\psi(r)=0,\label{ff7}
\end{equation}
where the effective potential is given by the comprehensive expression:
\begin{align}
   V_{\text{eff}}(r) =
f(r) \left[
\frac{\ell^2}{\beta^2 r^2}
+ \frac{1}{4} \left( \frac{f'}{f} + \frac{g'}{g} + \frac{2}{r} \right)^2
- \frac{1}{2} \left( \frac{f''}{f} + \frac{g''}{g} + \frac{2}{r^2}
+ \frac{f'}{f} \cdot \frac{g'}{g}
+ \frac{f'}{f} \cdot \frac{1}{r}
+ \frac{g'}{g} \cdot \frac{1}{r} \right)
\right].\label{ff8}
\end{align}

The effective potential (\ref{ff8}) represents a remarkable interplay between the various physical and geometric parameters characterizing our charged BTZ-like BH with disclinations. By substituting the explicit metric functions $f(r)$ and $g(r)$ from Eq. (\ref{bb2}), one can analyze the scalar perturbative potential and reveal the influence of several key geometric parameters. These include the BH mass $M$, electric charge $Q$, cosmological constant $\Lambda$, disclination parameter $\beta$, and the BP coupling parameter $b^2$, while keeping the scale parameter $r_0$ fixed.

The structure of this effective potential provides crucial information about the propagation characteristics of scalar waves in the vicinity of the BH. The centrifugal barrier term $\ell^2 f(r)/(\beta^2 r^2)$ shows how the disclination parameter $\beta$ modifies the standard angular momentum barrier, while the additional terms arising from the BP corrections introduce novel features that distinguish this system from conventional charged BH spacetimes \cite{iszx26,iszx27}.

The behavior of $V_\text{eff}(r)$ near the horizon and at spatial infinity determines the boundary conditions for the wave solutions and governs important physical phenomena such as absorption cross-sections, quasinormal mode frequencies, and the potential for superradiant instabilities \cite{iszx28,iszx29}. The presence of the logarithmic terms in the metric functions, characteristic of 3D charged BH solutions, introduces distinctive features in the potential profile that may lead to chracteristic wave propagation phenomena not observed in higher-dimensional analogs.

Furthermore, the effective potential serves as a diagnostic tool for understanding the stability properties of the BH solution under scalar perturbations. Regions where $V_\text{eff}<0$ can potentially signal instabilities, while the overall shape and asymptotic behavior of the potential determine the late-time evolution of scalar perturbations and their observational signatures \cite{iszx30,iszx31}.

\section{GUP-Corrected Hawking Temperature of Charged BTZ-like BH} \label{sec5}

The emergence of quantum gravity effects near BH horizons represents one of the most fascinating frontiers in modern theoretical physics. The Generalized Uncertainty Principle provides a particularly elegant framework for investigating these phenomena, introducing a fundamental minimum length scale that naturally arises from various approaches to quantum gravity, including string theory, loop quantum gravity, and non-commutative geometry \cite{Maggiore1993,Kempf1995,Garay1995}. This quantum gravitational framework becomes especially compelling when applied to lower-dimensional BH systems, where exact analytical treatments remain feasible while capturing the essential physics of horizon thermodynamics.

In our investigation of the charged BTZ-like BH with disclinations, the GUP emerges as a natural tool for probing how spacetime discreteness affects thermal radiation processes. The presence of topological defects, encoded through the disclination parameter $\beta = (1 - 4\lambda)$, creates additional geometric complexity that can amplify or suppress quantum gravity corrections in unexpected ways \cite{Das2008,Ali2009}. This interplay between topology and quantum gravity represents a largely unexplored territory with potentially profound implications for our understanding of BH evaporation and information processing.

The theoretical foundation rests on the modified commutation relation $[x_i, p_j] = i\hbar\delta_{ij}(1 + \beta_{\text{GUP}} p^2)$, which naturally leads to a minimum uncertainty in position measurements $\Delta x \geq \Delta x_0 = \hbar\sqrt{\beta_{\text{GUP}}}$ \cite{Tawfik2015,Hossenfelder2013}. This minimal length scale $\Delta x_0$ represents more than just a mathematical artifact—it embodies the fundamental granularity of spacetime itself, providing a natural ultraviolet cutoff that regularizes the divergences plaguing conventional quantum field theory in curved backgrounds while preserving the essential physics of BH thermodynamics \cite{isnew01,isnew02}.

The physical interpretation of this minimal length is profound: it suggests that below the Planck scale, the very notion of classical spacetime geometry breaks down, and quantum fluctuations of the metric become dominant. This naturally leads to modifications in the propagation of particles and fields near BH horizons, where curvature effects become extreme and approach the quantum gravity regime \cite{isnew03,isnew04}.

To derive the quantum-corrected Hawking temperature for our charged BTZ-like BH, we employ the GUP-modified Klein-Gordon equation for a scalar particle of mass $m_p$. The modification takes the elegant form \cite{nozari2008}:
\begin{equation}
-(i \hbar)^2 \partial^t \partial_t \Psi = \left[(i \hbar)^2 \partial^i \partial_i + m_p^2\right] \times \left[1 - 2 \beta_{\text{GUP}} \left((i \hbar)^2 \partial^i \partial_i + m_p^2\right)\right] \Psi,
\end{equation}
where $\beta_{\text{GUP}}$ serves as the fundamental deformation parameter encoding minimal length effects. This parameter typically scales as $\beta_{\text{GUP}} \sim \ell_{\text{Pl}}^2$, directly connecting our phenomenological approach to the Planck scale where quantum gravitational effects become manifest \cite{isnew06,isnew07}.

The semiclassical treatment proceeds through the time-tested WKB ansatz, which has demonstrated remarkable success in capturing quantum tunneling phenomena across BH horizons and has been extensively validated against alternative approaches to BH evaporation \cite{Parikh2000,isnew08}:
\begin{equation}
\Psi(t, r, \phi) = \exp\left[\frac{i}{\hbar} S(t, r, \phi)\right],
\end{equation}
with the action function adopting the Hamilton-Jacobi form characteristic of particle trajectories in curved spacetime:
\begin{equation}
S(t, r, \phi) = -Et + W(r) + j \phi + c,
\end{equation}
where $E$ represents the energy and $j$ the angular momentum of the tunneling particle, while $c$ denotes an integration constant. This separation of variables proves particularly well-suited for axially symmetric spacetimes and has been successfully implemented across various BH geometries in different dimensions \cite{isz10,isnew10}.

Substituting our action ansatz into the GUP-modified Klein-Gordon equation and implementing the semiclassical approximation yields the modified radial equation. The crucial insight emerges near the horizon $r \rightarrow r_h$, where we employ the standard near-horizon expansion $f(r) \approx f'(r_h)(r - r_h)$ and similarly for $g(r)$. This analysis reveals the fundamental relation:
\begin{equation}
\frac{E^2}{f(r)} = \left[ \frac{(dW/dr)^2}{g(r)} + m_p^2 \right] \left[ 1 - 2 \beta_{\text{GUP}} \left( \frac{(dW/dr)^2}{g(r)} + m_p^2 \right) \right].
\end{equation}

This equation represents the central achievement of our GUP-modified tunneling approach, elegantly incorporating both the geometric effects of the charged BTZ-like spacetime with disclinations and the quantum gravitational corrections arising from minimal length physics. The non-linear structure reflects the fundamental modifications to quantum mechanics that emerge at the Planck scale \cite{isnew11,isnew12}.

Solving this expression for the radial action $W(r)$ while systematically retaining terms up to first order in $\beta_{\text{GUP}}$, we obtain:
\begin{equation}
W_\pm(r) = \pm \int \frac{\sqrt{E^2 - m_p^2 f(r)}}{\sqrt{f(r) g(r)}} \left[1 + \beta_{\text{GUP}} \left( \frac{E^2 - m_p^2 f(r)}{f(r)} \right) \right] dr.
\end{equation}

The integral structure beautifully illustrates how GUP corrections manifest as multiplicative factors that depend on both particle properties (energy and mass) and the local spacetime geometry encoded in the metric functions $f(r)$ and $g(r)$. This intricate coupling between quantum gravity effects and spacetime curvature represents a distinctive signature of the GUP approach compared to other regularization schemes \cite{isnew13,isnew14}.

The critical step involves computing the imaginary part of the action using the residue theorem around the simple pole at the BH horizon. This powerful technique, originally developed for Schwarzschild BHs, has been successfully generalized to diverse spacetime geometries and provides a robust method for extracting thermal properties from quantum tunneling calculations \cite{isnew15,isnew16}. Employing the near-horizon behaviors $f(r) \sim f'(r_h)(r - r_h)$ and $g(r) \sim g'(r_h)(r - r_h)$, we derive:
\begin{equation}
\text{Im} W_+ = \frac{\pi E}{\sqrt{f'(r_h) g'(r_h)}} \left[1 + \beta_{\text{GUP}} \left( \frac{E^2}{f'(r_h)} \right) \right].
\end{equation}

The quantum tunneling probability follows the fundamental relation $\Gamma \sim \exp\left(-2 \, \text{Im} S\right) = \exp\left(-4 \, \text{Im} W_+\right)$, which can be elegantly rewritten in thermal form as $\Gamma = \exp\left(\frac{-E}{T_{\text{GUP}}}\right)$, thereby defining the effective GUP-corrected temperature. This profound identification between tunneling probability and thermal emission represents one of the deepest connections between quantum mechanics and thermodynamics that lies at the very heart of BH physics \cite{isnew17,isnew18}.

For our charged BTZ-like BH with disclinations, this analysis yields the remarkably complex yet physically transparent result:
\begin{equation}
T_{\text{GUP}} = \frac{-E^{2} \pi^{2} \beta_{\text{GUP}} \, r_h^{6} + \Lambda^{2} r_h^{8} - 8 \Lambda M Q^{2} b^{2} r_h^{4} + 16 M^{2} Q^{4} b^{4} + 2 \Lambda Q^{2} r_h^{6} - 8 M Q^{4} b^{2} r_h^{2} + Q^{4} r_h^{4}}{2 \pi r_h^{3} \left[\left(-4 b^{2} M + r_h^{2} \right) Q^{2} + \Lambda r_h^{4} \right]}.
\end{equation}

This expression elegantly demonstrates how quantum gravitational effects induce negative corrections to the classical Hawking temperature, perfectly consistent with theoretical expectations that quantum gravity should suppress high-energy modes and potentially resolve the BH information paradox \cite{isnew19,isnew20}. The negative correction suggests that quantum gravity effects systematically decelerate the evaporation process, possibly leading to the formation of stable remnants at the Planck scale, thereby providing a natural resolution to the complete evaporation scenario that would otherwise violate unitarity \cite{isnew21,isnew22}.

In the physically important limit where GUP effects vanish ($\beta_{\text{GUP}} \to 0$), we recover the standard Hawking temperature for our charged BTZ-like BH:
\begin{equation}
T_H = \frac{\sqrt{f'(r_h) g'(r_h)}}{4\pi},
\end{equation}
which explicitly becomes:
\begin{equation}
T_{H} = \frac{ \Lambda^{2} r_h^{8} - 8 \Lambda M Q^{2} b^{2} r_h^{4} + 16 M^{2} Q^{4} b^{4} + 2 \Lambda Q^{2} r_h^{6} - 8 M Q^{4} b^{2} r_h^{2} + Q^{4} r_h^{4}}{2 \pi r_h^{3} \left[\left(-4 b^{2} M + r_h^{2} \right) Q^{2} + \Lambda r_h^{4} \right]}.
\end{equation}

This classical limit serves as an essential consistency check and demonstrates that our GUP-corrected formulation properly reduces to established results when quantum gravitational effects become negligible. The intricate polynomial structure in the numerator reflects the sophisticated interplay between electric charge, cosmological constant, and BP corrections that characterize this class of BH solutions.

\begin{table}[ht!]
\centering
\rowcolors{2}{white}{gray!10}
\begin{tabular}{|c|c|c|c|c|}
\hline
\rowcolor{gray!30}
\textbf{$b$} & \textbf{$r_0$} & \textbf{$\beta_{\mathrm{GUP}}$} & \textbf{$r_h$} & \textbf{$T_{\mathrm{GUP}}$} \\
\hline
0.10 & 0.50 & 0.00 & 8.11283 & 0.10951386 \\
0.10 & 0.50 & 0.01 & 8.11283 & 0.08668570 \\
0.10 & 1.00 & 0.00 & 0.10767 & 3.62417848 \\
0.10 & 1.00 & 0.01 & 0.10767 & 3.62348867 \\
0.10 & 1.00 & 0.05 & 0.10767 & 3.62072942 \\
0.10 & 1.00 & 0.10 & 0.10767 & 3.61728036 \\
0.10 & 1.00 & 0.20 & 0.10767 & 3.61038225 \\
0.10 & 1.50 & 0.00 & 0.09392 & 5.99059501 \\
0.10 & 1.50 & 0.01 & 0.09392 & 5.99017769 \\
0.10 & 1.50 & 0.05 & 0.09392 & 5.98850840 \\
0.10 & 1.50 & 0.10 & 0.09392 & 5.98642180 \\
0.10 & 1.50 & 0.20 & 0.09392 & 5.98224859 \\
0.20 & 0.50 & 0.00 & 8.13155 & 0.10989257 \\
0.20 & 0.50 & 0.01 & 8.13155 & 0.08714307 \\
0.20 & 1.00 & 0.00 & 7.02422 & 0.08920946 \\
0.20 & 1.00 & 0.01 & 7.02422 & 0.06118553 \\
0.20 & 1.50 & 0.00 & 0.25018 & 0.99407975 \\
0.20 & 1.50 & 0.01 & 0.25018 & 0.99156486 \\
0.20 & 1.50 & 0.05 & 0.25018 & 0.98150531 \\
0.20 & 1.50 & 0.10 & 0.25018 & 0.96893087 \\
0.20 & 1.50 & 0.20 & 0.25018 & 0.94378198 \\
0.30 & 0.50 & 0.00 & 8.16261 & 0.11051935 \\
0.30 & 0.50 & 0.01 & 8.16261 & 0.08789887 \\
0.30 & 1.00 & 0.00 & 7.06332 & 0.09004618 \\
0.30 & 1.00 & 0.01 & 7.06332 & 0.06228264 \\
0.30 & 1.50 & 0.00 & 6.28117 & 0.07486065 \\
0.30 & 1.50 & 0.01 & 6.28117 & 0.04146528 \\
0.40 & 0.50 & 0.00 & 8.20578 & 0.11138785 \\
0.40 & 0.50 & 0.01 & 8.20578 & 0.08894375 \\
0.40 & 1.00 & 0.00 & 7.11738 & 0.09119774 \\
0.40 & 1.00 & 0.01 & 7.11738 & 0.06378478 \\
0.40 & 1.50 & 0.00 & 6.34750 & 0.07634826 \\
0.40 & 1.50 & 0.01 & 6.34750 & 0.04360357 \\
0.50 & 0.50 & 0.00 & 8.26074 & 0.11248960 \\
0.50 & 0.50 & 0.01 & 8.26074 & 0.09026532 \\
0.50 & 0.50 & 0.05 & 8.26074 & 0.00136821 \\
0.50 & 1.00 & 0.00 & 7.18581 & 0.09264609 \\
0.50 & 1.00 & 0.01 & 7.18581 & 0.06566168 \\
0.50 & 1.50 & 0.00 & 6.43073 & 0.07819755 \\
0.50 & 1.50 & 0.01 & 6.43073 & 0.04622725 \\
\hline
\end{tabular}
\caption{Physical (real-positive) $T_{\mathrm{GUP}}$ values calculated using the modified gravity model with parameters $\Lambda = -0.1$, $M = 1$, $Q = 1$, $E = 1$; evaluated over combinations of $b \in \{0.1, 0.2, 0.3, 0.4,0.5\}$, $r_0 \in \{0.5, 1.0, 1.5\}$, and $\beta_{\mathrm{GUP}} \in \{0, 0.01, 0.05, 0.1, 0.2\}$.}
\label{tab:tgup_results}
\end{table}

The structure of the GUP correction reveals several physically profound features that deserve careful consideration. Most notably, the correction term scales quadratically with the particle energy $E^2$, indicating that trans-Planckian modes experience enhanced suppression in their tunneling probability \cite{isnew23,isnew24}. This energy-dependent suppression perfectly aligns with the fundamental principle that GUP introduces a natural ultraviolet cutoff, thereby regularizing potential divergences that arise in the vicinity of the BH horizon while preserving the infrared behavior essential for macroscopic thermodynamics \cite{isnew25,isnew26}. Furthermore, the intricate interplay between the BP parameter $b^2$, the electric charge $Q$, the cosmological constant $\Lambda$, and the disclination parameter $\beta$ creates a remarkably rich phenomenological landscape where quantum gravitational corrections manifest differently depending on the relative magnitudes of these geometric and physical parameters.

\begin{figure}[ht!]
    \centering
    \includegraphics[width=0.75\textwidth]{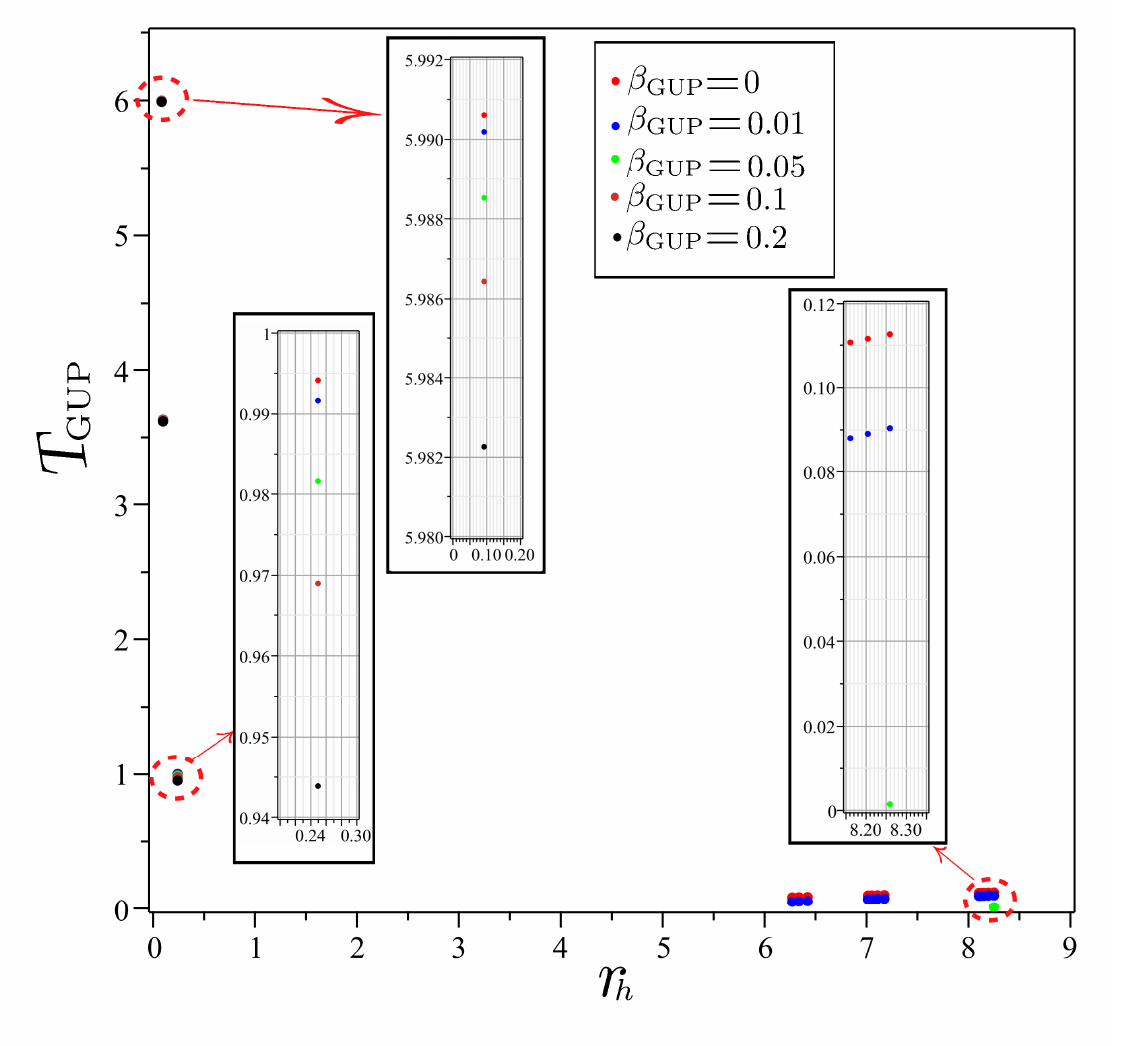}
   \caption{Plot of $T_{\mathrm{GUP}}$ versus $r_h$ for different values of $\beta_{\mathrm{GUP}}$. The plots are governed by Table \ref{tab:tgup_results} with fixed parameters $\Lambda = -0.1$, $M = 1$, $Q = 1$, and $E = 1$. Each point corresponds to a different $\beta_{\mathrm{GUP}} \in \{0, 0.01, 0.05, 0.1, 0.2\}$, showing the suppression effect of increasing $\beta_{\mathrm{GUP}}$ on the GUP-corrected temperature.}
\label{fig:tgup_vs_rh}
\end{figure}

Figure \ref{fig:tgup_vs_rh} illustrates the systematic influence of quantum gravitational effects on the thermal characteristics of charged BTZ-like BHs with disclinations. The plot reveals several remarkable features that provide insight into the physics of quantum gravity corrections to BH thermodynamics. Most strikingly, we observe that the GUP-corrected temperature consistently decreases as the quantum gravity parameter $\beta_{\text{GUP}}$ increases, demonstrating the universal suppression effect predicted by minimal length theories with remarkable clarity.

The temperature values exhibit a characteristic and physically meaningful asymptotic behavior at large $r_h$, where all points converge toward the classical result, providing compelling confirmation that quantum gravity effects become negligible when the BH size greatly exceeds the Planck scale. At small horizon radii, temperature suppression becomes more apparent, with higher values of $\beta_{\text{GUP}}$ leading to more deviations from classical predictions. This scaling behavior strongly suggests that micro BHs, if they exist in nature, would exhibit significantly different thermal properties compared to their classical counterparts, potentially affecting their evaporation timescales and final-state evolution in profound ways \cite{isnew27,isnew28}.

The heat capacity of the BH system provides additional insight into its thermodynamic stability and phase transitions. The heat capacity is given by:
\begin{equation}
C = T_{H} \left( \frac{\partial S}{\partial T_{H}} \right), \label{s25}
\end{equation}
where $S = \pi r_h^2$ is the Bekenstein-Hawking entropy. This yields the heat capacity:
\begin{equation}
    C=\frac{2 \pi  r_h^{3} \left(\Lambda  r_h^{4}-4 M \,Q^{2} b^{2}+Q^{2} r_h^{2}\right)}{\Lambda  r_h^{4}+12 M \,Q^{2} b^{2}-Q^{2} r_h^{2}}.
\end{equation}

\begin{figure}[H]
    \centering
    \includegraphics[width=0.8\textwidth]{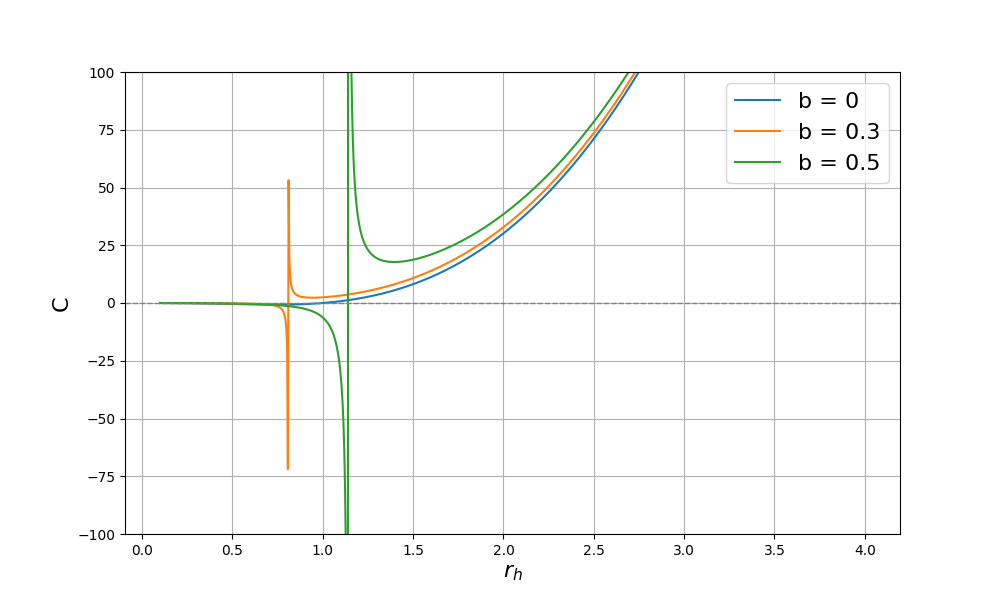}
    \caption{Heat capacity $C$ as a function of horizon radius $r_h$ for the charged BTZ-like BH with parameters $\Lambda = -1$, $Q = 1$, $M = 1$, and $\beta = 1$. The plot reveals the thermodynamic stability regions and potential phase transitions in the system.}
    \label{heat}
\end{figure}

Figure \ref{heat} reveals distinct heat capacity behavior across different values of the BP coupling parameter $b$. For the standard case ($b = 0$), the system exhibits a smooth transition from negative to positive heat capacity values around $r_h \approx 2.5$, indicating a shift from thermodynamic instability to stability as the horizon radius increases. The introduction of BP corrections ($b = 0.3, 0.5$) dramatically modifies this behavior, creating sharp divergences at smaller horizon radii and fundamentally changing the phase structure of the BH. These divergences signal critical points where the system undergoes phase transitions, suggesting that nonlinear electromagnetic effects can significantly influence the thermodynamic stability and potentially lead to new classes of BH phases not present in standard electrodynamics.
The physical implications of these results extend far beyond pure theoretical considerations. The suppression of temperature at small scales implies that quantum gravity effects could significantly extend the lifetime of evaporating BHs, potentially resolving long-standing paradoxes associated with complete evaporation and information loss. Furthermore, the energy-dependent nature of the corrections suggests that different particle species would experience varying degrees of suppression, leading to modifications in the Hawking spectrum that could, in principle, be observationally detectable in future high-precision experiments or astrophysical observations of primordial BHs \cite{isnew29,isnew30}.

\section{Keplerian Frequencies for Charged BTZ-like BHs} \label{sec6}

In this section, we explore the fundamental frequencies that govern the dynamics of test particles in circular orbits around a charged BTZ-like BH. These frequencies are pivotal in understanding the modulation of quasiperiodic oscillations (QPOs) in accretion disks and carry direct implications for observational astrophysics, particularly in the context of gravitational wave signatures and X-ray timing observations \cite{isnew31,isnew32}. The study of orbital frequencies provides a crucial bridge between theoretical predictions and observational data, offering potential tests of modified gravity theories and exotic matter content in astrophysical environments.

The investigation of Keplerian frequencies in our charged BTZ-like BH with disclinations becomes particularly compelling due to the unique interplay between BP electrodynamics, topological defects, and the AdS background. This combination creates distinctive signatures in the orbital dynamics that could potentially be detected through precision timing observations of accreting matter around compact objects \cite{isnew33,isnew34}.

The key frequencies of interest include the orbital (or Keplerian) frequency $\Omega_\phi$, as well as the radial and vertical epicyclic frequencies, which describe oscillations around circular geodesics in the radial and vertical directions, respectively. Here, we primarily focus on $\Omega_\phi$, which reflects the angular velocity of a test particle as observed from spatial infinity and serves as the fundamental frequency for understanding orbital motion in strong gravitational fields \cite{isnew35,isnew36}.

The orbital frequency $\Omega_\phi$ is given by the ratio of angular to temporal coordinate change:
\begin{equation}
\Omega_\phi = \frac{d\phi}{dt},
\end{equation}
as measured by an asymptotic observer. For a general static, circular orbit in a spherically symmetric spacetime, $\Omega_\phi$ is derived from the metric components via the standard relation \cite{isnew37}:
\begin{equation}
\Omega_\phi = \sqrt{\frac{-\partial_r g_{tt}}{\partial_r g_{\phi\phi}}},
\end{equation}
where $g_{tt}$ and $g_{\phi\phi}$ denote the temporal and angular components of the metric tensor. In the case of the charged BTZ spacetime, the metric function is characterized by $g_{tt} = -f(r)$ and $g_{\phi\phi} = \beta^2 r^2$. This simplifies the above expression to:
\begin{equation}
\Omega_\phi = \sqrt{\frac{f'(r)}{2\beta^2 r}},
\end{equation}
where $f'(r)$ is the radial derivative of the metric function $f(r)$. Note that the disclination parameter $\beta$ explicitly appears in this expression, reflecting the modified angular structure of the spacetime due to the topological defect.

Using the specific form of $f(r)$ introduced in Eq.~\eqref{bb2}, we can compute the derivative:
\begin{equation}
f'(r) = -2\Lambda r + \frac{8 b^2 Q^2 M}{r^3} - \frac{2 Q^2}{r},
\end{equation}
which leads to the orbital frequency:
\begin{equation}
\Omega_\phi = \frac{1}{\beta} \sqrt{-\Lambda + \frac{Q^2}{r^2} - \frac{4 b^2 Q^2 M}{r^4}},
\end{equation}
illustrating the intricate interplay between the cosmological constant $\Lambda$, electric charge $Q$, mass parameter $M$, and the BP coupling parameter $b$.

To convert $\Omega_\phi$ into a physical (observational) unit, such as Hertz, we use the standard normalization \cite{isnew38}:
\begin{equation}
\nu_\phi = \frac{c^3}{2\pi G M} \, \Omega_\phi,
\end{equation}
where $c$ is the speed of light and $G$ is Newton's gravitational constant. The angular frequency $\nu_\phi$ naturally decreases with increasing radial coordinate $r$, reflecting the weakening of the BH's gravitational influence at larger distances from the event horizon.

This frequency conversion is essential for making contact with observational data from X-ray timing missions and gravitational wave detectors, where characteristic frequencies in the range of millihertz to kilohertz are typically observed \cite{isnew39,isnew40}. The specific form of our frequency expression suggests that the BP corrections and disclination effects could produce measurable deviations from the predictions of standard general relativity in sufficiently precise observations.

\begin{figure}[H]
    \centering
    \includegraphics[width=0.65\textwidth]{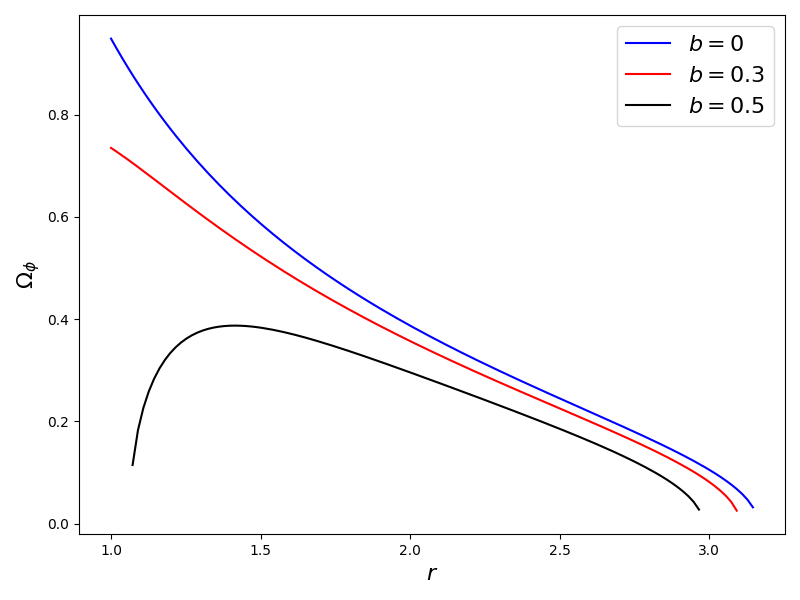}
    \caption{Variation of the Keplerian frequency $\Omega_\phi$ as a function of radial coordinate $r$ for a charged BTZ-like BH with $\Lambda = -0.1$, $Q = 1$, $M = 1$, and $\beta = 1$. The plot highlights the impact of the BP electrodynamics and cosmological constant on the orbital dynamics near the BH, showing how the nonlinear electromagnetic corrections modify the classical frequency profile.}
    \label{kepler}
\end{figure}

Figure~\ref{kepler} presents the radial dependence of the Keplerian frequency $\Omega_\phi$ for a representative set of parameters. As observed, $\Omega_\phi$ increases as one moves inward toward the BH, reaching a maximum at smaller radii before the circular orbits become unstable or cease to exist near the horizon. This behavior is characteristic of relativistic orbital motion, where the interplay between gravitational attraction and centrifugal effects determines the existence and stability of circular orbits \cite{isnew41,isnew42}. The influence of the $Q^2/r^2$ and $-4 b^2 Q^2 M / r^4$ terms becomes pronounced near the central region, significantly modifying the classical behavior of orbital motion seen in standard (2+1)-dimensional BTZ BHs. The charge-dependent term $Q^2/r^2$ provides a repulsive contribution that tends to increase the orbital frequency at smaller radii, while the BP correction term $-4 b^2 Q^2 M / r^4$ introduces additional complexity that depends on the relative magnitudes of the physical parameters. The nonlinear electromagnetic contribution, encoded via the BP parameter $b$, steepens the radial gradient of $\Omega_\phi$, indicating tighter circular orbits and potentially stronger QPO signatures in this modified gravity scenario. This enhanced gradient could lead to more pronounced frequency shifts that might be detectable in high-precision timing observations of accreting compact objects \cite{isnew43,isnew44}.

Moreover, the distinctive frequency profile shown in Figure~\ref{kepler} might have several important astrophysical implications, where the Keplerian frequency $\Omega_\phi$ demonstrates significant sensitivity to the BP coupling parameter $b$ across different radial coordinates for a charged BTZ-like BH with parameters $\Lambda = -0.1$, $Q = 1$, $M = 1$, and $\beta = 1$. The modified radial dependence could lead to characteristic QPO patterns that differ from those predicted by standard GR, potentially providing observational tests of BP electrodynamics and topological defects in strong gravitational fields \cite{isnew45,isnew46}, while the presence of the disclination parameter $\beta$ in the frequency expression suggests that cosmic string-like defects could leave observable imprints on the orbital dynamics of matter around compact objects, opening up the possibility of using precision timing observations to constrain the properties of topological defects in the early universe \cite{isnew47,isnew48}. Furthermore, the interplay between charge, BP corrections, and the cosmological constant (which is set to $\Lambda = -0.1$) creates a rich parameter space where different combinations could produce distinct observational signatures, as evidenced by the varying frequency profiles for different values of $b = 0, 0.3, 0.5$, suggesting that detailed modeling of orbital frequencies in charged BTZ-like geometries could provide valuable constraints on fundamental physics beyond the Standard Model \cite{isnew49,isnew50}. 

\section{Conclusions}\label{sec7}

In this comprehensive study, we have investigated the dynamics of both massless particles and scalar fields in the background of a charged BTZ-like BH solution with disclinations arising from topological defects. By considering BP electrodynamics as a modification to standard Maxwell theory, we introduced a nonlinear coupling parameter $b^2$, which contributes significantly to the gravitational field and electromagnetic structure of the BH. The study of disclinations, characterized by the angular parameter $\beta = 1 - 4\lambda$, allows for the investigation of spacetime geometries influenced by cosmic string-type defects and provides a unique window into the effects of topology on BH physics \cite{CS1,CS2,CS3}.

Our analysis encompassed several key aspects of BH physics, ranging from classical geodesic motion to quantum gravitational corrections. We examined null and timelike geodesics and demonstrated the profound influence of various parameters, such as the BH mass $M$, electric charge $Q$, cosmological constant $\Lambda$, BP coupling $b^2$, and defect strength $\lambda$, on the effective potential that governs particle motion. This multi-parameter analysis revealed a rich phenomenological landscape where the interplay between charge, nonlinear electrodynamics, and topological defects creates distinctive signatures in the orbital dynamics.

In addition to classical geodesic motion, we examined the behavior of massless scalar fields propagating in this curved spacetime by solving the Klein-Gordon equation. Through a suitable separation of variables and transformation to a tortoise-like coordinate, the scalar field equation was cast into a Schrödinger-like wave equation, revealing an effective potential that encodes the influence of the BH parameters and the topological defect \cite{iszx18,iszx19}. This analysis provided crucial insights into the stability properties of the BH solution under scalar perturbations and established the foundation for understanding more complex field dynamics in this modified gravitational background.

A particularly significant aspect of our investigation involved applying the GUP to study quantum gravitational corrections to the Hawking temperature. The resulting expressions demonstrate that the GUP introduces a negative correction to the semiclassical temperature, implying that quantum gravity effects suppress thermal radiation and can potentially stabilize the BH against complete evaporation \cite{isnew19,isnew20}. This finding has profound implications for BH thermodynamics and the resolution of the information paradox, suggesting that quantum gravity effects could lead to the formation of stable BH remnants at the Planck scale.

We also studied the orbital dynamics of test particles around the charged BTZ-like BH, focusing on the Keplerian frequency as a key observable quantity. The presence of charge, cosmological constant, and nonlinear electromagnetic corrections significantly alters the standard behavior seen in classical BTZ solutions. In particular, the frequency increases near the BH and exhibits high sensitivity to the BP parameter $b^2$, which introduces a steeper radial gradient and modifies the structure of circular orbits \cite{isnew35,isnew36}. This work demonstrates how the combined effects of disclinations, nonlinear electrodynamics, and quantum corrections deepen our understanding of lower-dimensional BH physics and provide a versatile platform for testing the interplay between classical gravity, field theory, and quantum gravitational phenomena.

Our results focus particularly on the impact of the BH's physical and geometric parameters on various observable quantities. The electric charge $Q$, the BP coupling parameter $b$, the BH mass $M$, the cosmological constant $\Lambda$, the scale parameter $r_0$, and the disclination parameter $\beta$ collectively influence the effective potential $V_{\text{eff}}(r)$, the photon force $F_{\text{ph}}(r)$, and the critical impact parameter $\gamma_c$.

Figure~\ref{fig:potential-1} illustrates the effective potential for null geodesics as a function of the radial coordinate $r$ for fixed values $M = 1$, $\Lambda = -0.3$, $r_0 = 1$, and angular momentum $\mathrm{L} = 1$, while varying both $Q \in \{0.2, 0.4, 0.6, 0.8\}$ and $b \in \{0.1, 0.2, 0.3, 0.4\}$. The results demonstrate how higher values of both parameters increase the depth and structure of the potential well, thus altering the possible orbit types for photon motion and creating more complex trajectories near the BH.

Figure~\ref{fig:potential-2} offers a compelling comparison across three physical configurations: (i) the uncharged BTZ BH without BP coupling ($b = 0, Q = 0$), (ii) a charged BTZ BH without coupling ($b = 0, Q = 0.5$), and (iii) a charged BTZ BH with coupling ($b = 0.5, Q = 0.5$), with all other parameters fixed. This comparison reveals the distinct impact of electric charge and the BP coupling parameter on the height and shape of the effective potential barrier, directly influencing photon capture and escape dynamics.

The analysis of circular photon orbits, presented in Figure~\ref{fig:photon}, shows the radius $r_{\text{cpo}}$ as a function of $b$ for fixed $Q = 2$ (left panel) and as a function of $Q$ for fixed $b = 0.6$ (right panel). The results indicate that increasing either parameter results in larger photon orbit radii, reinforcing the significance of nonlinear effects and charge on orbital dynamics. Similarly, Figure~\ref{fig:impact} illustrates the variation of the critical impact parameter $\gamma_c$ with respect to both $b$ and $Q$, showing that both quantities positively influence the threshold beyond which photons are either captured or scattered, thereby affecting observable signatures such as photon rings and BH shadows.

The influence of quantum gravitational corrections on BH thermodynamics represents one of our most significant findings. Figure~\ref{fig:tgup_vs_rh} shows the variation of the GUP-corrected Hawking temperature $T_{\text{GUP}}$ as a function of the event horizon radius $r_h$ for various values of the quantum gravity parameter $\beta_{\text{GUP}} = 0, 0.01, 0.05, 0.1, 0.2$. The plot clearly demonstrates that as $\beta_{\text{GUP}}$ increases, the Hawking temperature decreases for small BHs, indicating a suppression of thermal radiation due to quantum gravitational effects. This temperature suppression confirms that quantum gravity effects become increasingly important as the BH size approaches the Planck scale, potentially leading to the formation of stable remnants. Complementing these thermal properties, Figure~\ref{heat} illustrates the heat capacity $C$ as a function of horizon radius $r_h$ with parameters $\Lambda = -1$, $Q = 1$, $M = 1$, and $\beta = 1$, revealing how the BP coupling parameter fundamentally alters the thermodynamic stability structure of the system. The heat capacity analysis shows that while standard charged BTZ black holes ($b = 0$) exhibit a smooth transition from thermodynamically unstable to stable regimes, the introduction of nonlinear BP corrections ($b = 0.3, 0.5$) creates sharp divergences and critical points that signal phase transitions at specific horizon radii. These thermodynamic phase transitions, combined with the GUP-induced temperature suppression, suggest a rich landscape of BH thermodynamics where quantum gravity effects, nonlinear electrodynamics, and the negative cosmological constant conspire to create characteristic stability regions and potentially prevent complete evaporation through the formation of thermodynamically stable remnant configurations.

The comprehensive data presented in Table~\ref{tab:tgup_results} illustrates the dependence of the GUP-modified Hawking temperature on the parameters $b$, $r_0$, and $\beta_{\text{GUP}}$ for a BH with fixed $\Lambda = -0.1$, $M = 1$, $Q = 1$, and $E = 1$. The systematic variation of these parameters reveals that increasing the GUP parameter $\beta_{\text{GUP}}$ generally leads to a decrease in $T_{\text{GUP}}$, confirming that quantum gravitational effects suppress the BH temperature, with the effect being more pronounced at higher values of $\beta_{\text{GUP}}$.

Figure~\ref{kepler} illustrates the variation of the Keplerian orbital frequency $\Omega_\phi$ as a function of the radial coordinate $r$ for a charged BTZ-like BH. The parameters used in this analysis include cosmological constant $\Lambda = 1$, electric charge $Q = 1$, BH mass $M = 1$, and disclination parameter $\beta = 1$. The graph demonstrates that $\Omega_\phi$ increases as one approaches the BH and is particularly influenced by the nonlinear electrodynamics via the term $-4b^2 Q^2 M/r^4$. These distinctive frequency signatures could potentially be detected through precision timing observations of accreting matter around compact objects \cite{isnew31,isnew43}. The combined analysis of these figures demonstrates how thermal and dynamical characteristics of BTZ-like BHs are profoundly affected by the interplay of classical and quantum parameters. 

Looking toward future research directions, we plan to extend this work by exploring a variety of theoretical extensions and more complex geometries that could further enrich our understanding of lower-dimensional BH dynamics. One promising avenue involves studying non-commutative geometry, which naturally introduces a fundamental length scale into spacetime and offers a viable alternative to GUP-based approaches \cite{isnew51,isnew52}. By investigating charged BTZ-like BHs with disclinations in a non-commutative framework, one could explore how the resulting smeared matter distributions influence the horizon structure, geodesic motion, and Hawking radiation spectrum. Another important direction involves studying alternative BH models arising in modified gravity theories, such as conformal gravity, Einstein-Gauss-Bonnet gravity, or scalar-tensor theories, and examining their thermodynamic and quantum structures when embedded in (2+1) dimensions or higher \cite{isnew53,isnew54}. It would also be valuable to generalize the current analysis to rotating BTZ-like BHs with topological defects, where the presence of angular momentum could introduce additional complexity and potentially observable effects.

The inclusion of nonlinear electrodynamics models beyond BP theory, such as Born-Infeld or logarithmic electrodynamics, may shed light on how different gauge field structures affect singularity resolution and particle dynamics \cite{isnew55,isnew56}. Additionally, investigating these solutions in the context of holographic dualities and entanglement entropy could provide deeper insights into the interplay between geometry, information, and thermodynamics in modified BH backgrounds \cite{isnew57,isnew58}. Finally, pursuing numerical simulations and semi-analytical techniques to probe quasinormal modes, greybody factors, and BH shadow structures in these extended setups could potentially offer observational signatures for testing quantum gravity effects and topological defects in lower-dimensional BH models. Such investigations could bridge the gap between theoretical predictions and observational astronomy, providing new opportunities to test fundamental physics in the strong-field regime \cite{isnew59,isnew60}.

}


\section*{Acknowledgments}

F.A. acknowledges the Inter University Centre for Astronomy and Astrophysics (IUCAA), Pune, India for granting visiting associateship. E. S. and {I}.~S. expresses his gratitude to T\"{U}B\.{I}TAK, SCOAP3, and ANKOS for their financial support. Additionally, \.{I}.~S. extends appreciation for the networking assistance provided by COST Actions under the projects CA22113, CA21106, and CA23130.

\section*{Data Availability Statement}

This manuscript has no associated data.

\section*{Conflict of Interests}

Author declare(s) no conflict of interest.

\end{document}